\documentclass[fleqn,usenatbib]{mnras}

\usepackage{newtxtext,newtxmath}
\usepackage[T1]{fontenc}
\DeclareRobustCommand{\VAN}[3]{#2}
\let\VANthebibliography\thebibliography
\def\thebibliography{\DeclareRobustCommand{\VAN}[3]{##3}\VANthebibliography}

\usepackage{graphicx}
\usepackage{amsmath}

\usepackage{microtype}
\usepackage{enumerate}
\usepackage{xcolor}

\title[Bar lengths from morphological segmentation]{A morphological segmentation approach to determining bar lengths}

\author[M. K. Cavanagh et al.]{
Mitchell K. Cavanagh,$^{1}$\thanks{E-mail: mitchell.cavanagh@icrar.org (MKC)}
Kenji Bekki$^{1}$ and
Brent A. Groves$^{1,2}$
\\
$^{1}$International Centre for Radio Astronomy Research, The University of Western Australia, 7 Fairway, Crawley, WA 6009, Australia\\
$^{2}$Research School of Astronomy and Astrophysics, Australian National University, Mt Stromlo Observatory, Weston Creek, ACT 2611, Australia\\
}

\date{Accepted XXX. Received YYY; in original form ZZZ}

\pubyear{2023}

\begin{document}
\label{firstpage}
\pagerange{\pageref{firstpage}--\pageref{lastpage}}
\maketitle

\begin{abstract}
Bars are important drivers of galaxy evolution, influencing many physical processes and properties. Characterising bars is a difficult task, especially in large-scale surveys. In this work, we propose a novel morphological segmentation technique for determining bar lengths based on deep learning. We develop U-Nets capable of decomposing galaxy images into pixel masks highlighting the regions corresponding to bars and spiral arms. We demonstrate the versatility of this technique through applying our models to galaxy images from two different observational datasets with different source imagery, and to RGB colour and monochromatic galaxy imaging. We apply our models to analyse SDSS and Subaru HSC imaging of barred galaxies from the NA10 and SAMI catalogues in order to determine the dependence of bar length on stellar mass, morphology, redshift and the spin parameter proxy $\lambda_{R_e}$. Based on the predicted bar masks, we show that the relative bar scale length varies with morphology, with early type galaxies hosting longer bars. While bars are longer in more massive galaxies in absolute terms, relative to the galaxy disc they are actually shorter. We also find that the normalised bar length decreases with increasing redshift, with bars in early-type galaxies exhibiting the strongest rate of decline. We show that it is possible to distinguish spiral arms and bars in monochrome imaging, although for a given galaxy the estimated length in monochrome tends to be longer than in colour imaging. Our morphological segmentation technique can be efficiently applied to study bars in large-scale surveys and even in cosmological simulations.
\end{abstract}


\section{Introduction}

Stellar bars are centrally-located, rectangular-shaped morphological structures that have profound impacts on the physical, dynamical and morphological evolution of their host galaxies \citep{sellwood1993, abraham1999, masters2011, cheung2013, conselice2014}. Bars are known to influence gas dynamics and star formation \citep{shlosman1993, ellison2011, fanali2015, lin2020}, the redistribution of angular momentum \citep{weinberg1985, athanassoula2005}, disc-halo interactions \citep{athanassoula2002, valenzuela2003}, secular evolution and bulge growth \citep{rautiainen2002, kormendy2004, jogee2005, athanassoula2013book,kruk2018}, morphological transition and quenching \citep{masters2012,spinoso2017,fraser-mckelvie2020,geron2021}, and even AGN activity \citep{shlosman1989, alonso2014}. Bars are present in the majority of spiral galaxies. Observational estimates for the bar fraction range from 55\% to as high as 70\% of all spiral galaxies \citep{eskridge2000, aguerri2009, saha2018}, to around a third of all galaxies \citep{nair2010a}. It is well established that the prevalence of bars declines with increasing redshift \citep{sheth2008, cameron2010, melvin2014, kim2021}. The bar fraction is also known to vary with morphology and environment. Previous studies have found that bars are most common in redder galaxies such as early-type spirals \citep{combes1993, masters2011, skibba2012, vera2016, cervantessodi2017, erwin2018}, although other studies have asserted that there are just as prevalent, if not more, in late-type spirals \citep{erwin2018, tawfeek2022}. Bars are also known to be more common in dense environments \citep{elmegreen1990, marinova2009, lee2012}. Lenticular galaxies are also known to host bars, albeit in much lower fractions compared to spirals \citep{laurikainen2009}.

Many studies have focused on the prevalence of bars and the properties of barred galaxies, yet there is also much to be gleaned from studying the properties of the bars in their own right. In particular, the length of a galaxy bar is an important physical property that can provide important insights into the nature of its current evolution, as well as its effects on the host galaxy \citep{erwin2005, hoyle2011, erwin2019, fraser-mckelvie2020barlength, geron2021}. The length of a bar can be defined in terms of its actual, absolute length (in units of length), or as a dimensionless, normalised length with respect to the size of the galaxy, such as the scale radius of the host galaxy disc $R_d$. This study examines the bar length using both of these definitions. In particular, we will refer to the absolute bar length as $L_{\text{bar}}$, and the relative bar length as $L_{\text{bar}}/R_d$ where $R_d$ is the disc scale radius. Previous studies have used bar length as a means for measuring the ``strength'' of stellar bars \citep{aguerri1998, geron2021}. Furthermore, bar length is known to vary significantly with the morphology of the host galaxy. Studies based on both $N$-body simulations and observations of nearby barred galaxies have determined that bars are typically shorter in late-type galaxies \citep{elmegreen1985,combes1993,aguerri2009,erwin2019}, with \citet{erwin2005} finding that bars are over twice as long in early-type galaxies as opposed to late-type discs. One possible explanation for this is the higher gas fractions in late-type galaxies, which can suppress the growth of the bar \citep{bournaud2005, berentzen2007, athanassoula2013}. Bar lengths can serve as independent markers of the properties of the host galaxies. In particular, studies have also demonstrated the absolute bar length positively correlates with increasing galaxy mass \citep{kormendy1979, erwin2005, diaz-garcia2016}. Curiously, previous observational studies examining the evolution of barred galaxies have found little change in overall bar lengths with redshift \citep{sheth2008,kim2021}.

To measure the length of a bar, it is first necessary to confirm the existence of a bar through morphological classification. This can be achieved through visual classification, whether by groups of expert astronomers or crowdsourcing \citep{sheth2008, nair2010a, masters2011, buta2013,  masters2021}. There are also several, analytical methods for classifying barred galaxies, including isophotal ellipse fitting \citep{abraham1999, laine2002, erwin2005, menendez-delmestre2007, consolandi2016}, Fourier analysis (usually of the $m=0$ and $m=2$ modes of the azimuthal light profile) \citep{elmegreen1985, ohta1990, aguerri1998, odewahn2004,garcia-gomez2017}, and photometric structural decomposition \citep{reese2007, durbala2009, weinzirl2009}. Recent studies have also utilised non-parametric deep learning models to classify barred galaxies \citep{abraham2018, cavanagh2020, cavanagh2022}. Bar lengths are typically measured through analytical approaches, most notably isophotal ellipse fitting \citep{erwin2005, gadotti2006, marinova2007, hoyle2011} and photometric bulge/bar/disc decompositions \citep{gadotti2008, durbala2008}. However, these parametric approaches often rely on significant preprocessing and/or require auxiliary data in addition to galaxy imagery.

Over the last few years, deep learning techniques have increasingly been used across wide range of applications within astronomy, ranging from classification to regression to segmentation \citep{baron2019, fluke2020, huertas-company2023}. In this study, we propose a novel technique for estimating bar lengths via image segmentation with deep learning. We develop and train U-Nets: specialised convolutional neural networks capable of pixel-level image segmentation. Specifically, we train two U-Nets – one on RGB colour imaging, another on monochromatic imaging – to decompose an image of a galaxy into pixel masks highlighting the locations of spiral arms and stellar bars. It is then possible to estimate the length of bars through performing ellipse fitting on these pixel masks. Unlike parametric or analytical methods, such as isophotal ellipse fitting or Fourier analysis, that rely on a choice of free parameters or criteria in order to detect the presence of a bar, our U-Nets directly output a predicted bar mask solely from the image of a galaxy, without the need for auxiliary data. Only after the U-Net has extracted the bar mask from the image of a galaxy can arbitrary thresholds be applied to estimate the bar length with varying degrees of confidence.

Originally designed for biomedical imaging \citep{ronneberger2015}, U-Nets have been used in a broad range of applications including robotics, industrial automation, self-driving cars and mass surveillance (see \citealt{minaee2020} for a recent, comprehensive survey). Recently, studies in astronomy have adopted U-Nets for source detection, source deblending, image denoising and deconvolution, and even the structural decomposition of galaxy components \citep{boucaud2020, hausen2020, sureau2020, bekki2021, vojtekova2021, robertson2022}, however there is yet to be a study to specifically focus on stellar bars. With the U-Net models introduced in this study, it is possible to directly output pixel masks highlighting the regions corresponding to bars in barred galaxies simply through analysing galaxy images. Our U-Nets are trained using crowdsourced masks from the Galaxy Zoo 3D (GZ3D) dataset \citep{masters2021}. To demonstrate the versatility of these models, we apply these U-Nets to analyse images of galaxies across two datasets with different galaxy imaging: the \citet{nair2010} visual morphological catalogue (Sloan Digital Sky Survey RGB cutouts), and the Sydney AAO Multi-object IFS (SAMI) survey DR3 \citep{croom2021} (Subaru HyperSuprime Cam RGB cutouts; \citealt{aihara2022}).

The structure of our paper is as follows. In \S~\ref{s:methods}, we outline our methodology, briefly summarising the datasets used in the study before discussing our U-Net architecture, the data preprocessing and model training procedures, and the full bar length estimation pipeline, showcasing the sequence of steps from galaxy image to bar length. In \S~\ref{s:results}, we discuss our core results. We highlight the inherent versatility of our deep learning-based approach through directly applying the U-Net model to analyse images of barred galaxies from different observational datasets, each based on different imagery. We examine the predicted bar lengths with respect to various physical properties of galaxies, including stellar mass, morphology and kinematics, as well as redshift. In \S~\ref{s:discussion}, we discuss the implications of these results on the formation and evolution of bars, as well as the role they play in the evolution of their host galaxies. In particular, we examine the changes in the distribution of bar lengths over redshift, and how this relates to changes in galaxy size, including with respect to morphology. Also in \S~\ref{s:discussion}, we examine the differences between two U-Nets trained with colour imaging and monochrome imaging. In particular, we show that it is still possible for the monochrome U-Net to differentiate spiral arms from bars. We further discuss the utility and limitations of our novel deep learning approach, as well as further avenues for studying additional morphological subfeatures. Lastly, we summarise our key findings in \S~\ref{s:concl}.

\section{Methods}
\label{s:methods}

\subsection{Datasets}

This study utilises images of galaxies from multiple datasets. Of these, the most significant dataset is the GZ3D dataset \citep{masters2021} which is used to train the U-Net models. Other galaxy images include SDSS imaging for barred galaxies in the \citet{nair2010} morphological catalogue and HSC imaging of galaxies from the SAMI survey DR3 \citep{croom2021}. Here we briefly summarise these datasets.

\subsubsection{GZ3D}
\label{ss:gz3d}

The Galaxy Zoo 3D dataset \citep{masters2021} is a dataset of crowdsourced classification masks for 29,831 galaxies sourced from the initial Mapping Nearby Galaxies at APO (MaNGA) survey target list \citep{bundy2015, wake2017}. It is publicly available as a value-added catalogue in the SDSS Data Release 17 \citep{abdurrouf2022}. Participants were tasked with inspecting RGB colour galaxy images (SDSS cutouts) in a web-based interface and painting over region(s) corresponding to the desired feature. These features include core structural components such as spiral arms and galaxy bars, as well as the locations of foreground stars and the approximate location of the galaxy centre. The final masks for each galaxy are obtained by aggregating the individual masks drawn by each volunteer contributor. The resulting ``count'' masks are thus formatted such that the value of each pixel corresponds to the number of contributors that explicitly marked that pixel as belonging to the desired morphological feature.

For the purpose of this study, we are focused only on samples containing spiral masks and/or bar masks. The aim is for our U-Nets to output both a spiral and bar mask for a given image, rather than just a bar mask by itself. This is to force the model to distinguish spiral arms and bars as being distinct morphological features. The training data ultimately consists of the galaxy images as inputs, with the spiral and bar masks as the labels, or ground truth. We enforce a minimum confidence level by restricting our training data only to GZ3D galaxies whose bar or spiral masks have been drawn by at least three volunteer classifiers. This brings the total number of suitable galaxies down from the 29,831 total to 7,965 galaxies. We note that this does not completely preclude erroneous masks (as shown in Appendix~\ref{app:inputs}), and that the choice of the minimum number of classifiers is ultimately a trade-off between quantity and quality.

\subsubsection{NA10}

The \citet{nair2010} catalogue consists of morphological classifications for 14,034 galaxies sourced from the SDSS Data Release 4. The catalogue consists of galaxies brighter than 16 mag (extinction-corrected $g$-band) between $z \approx 0.01$ and $z = 1$. For the purpose of this work, our analysis is restricted to the 2,612 barred galaxies as sourced from the catalogue. We use SDSS \textit{gri} band colour images resized to $128\times 128$ pixels so that they can be input into our U-Net models. These cutouts were sourced from the latest data release of the SDSS via the publicly accessible \texttt{SkyServer} web interface.

\subsubsection{SAMI DR3}

The Sydney AAO Multi-object IFS (SAMI) survey \citep{croom2012, bryant2015, croom2021} is a large-scale integral-field spectroscopy survey of several thousand nearby, low-redshift galaxies. This includes their morphological classifications, which are described in \citet{cortese2016}, along with several key kinematic observables including the spin parameter proxy $\lambda_{R_e}$ \citep{brough2017, vandesande2021}. We first select 5,536 nearby, low-redshift galaxies from SAMI DR3 \citep{croom2021}, specifically all galaxies from the Galaxy and Mass Assembly (GAMA) survey DR3 input catalogue \citep{driver2011, bryant2015}. These image cutouts were initially 360$\times$360 pixels in size (corresponding to angular field of view of 60.48 arcseconds), however they were resized to 128x128 for use with the U-Net models. Although there are no explicit barred or unbarred galaxies in the SAMI DR3 catalogue, we apply an existing deep learning CNN, based on previous work, to classify all 5,536 galaxies as barred or unbarred. For the bar classification, we use $g$-band HyperSuprime Cam (HSC) \citep{aihara2022} images of all galaxies; this is since the bar classifier model is specifically trained to classify $g$-band imaging. For full details of this model, including its architecture and training, see \citet{cavanagh2022}. We henceforth only consider galaxies that are classified by this model as barred with a certainty greater than 50\%. For the purposes of bar extraction with the U-Net, we use the RGB cutouts of the galaxies classified by our bar model as barred. These RGB cutouts are publicly accessible as part of the HSC Public Data Release 3, courtesy of the NAOJ/HSC collaboration (see also \citealt{aihara2022}).

\subsection{The U-Net Model}
\label{ss:unetmodel}

\begin{figure*}
    \includegraphics[scale=0.255]{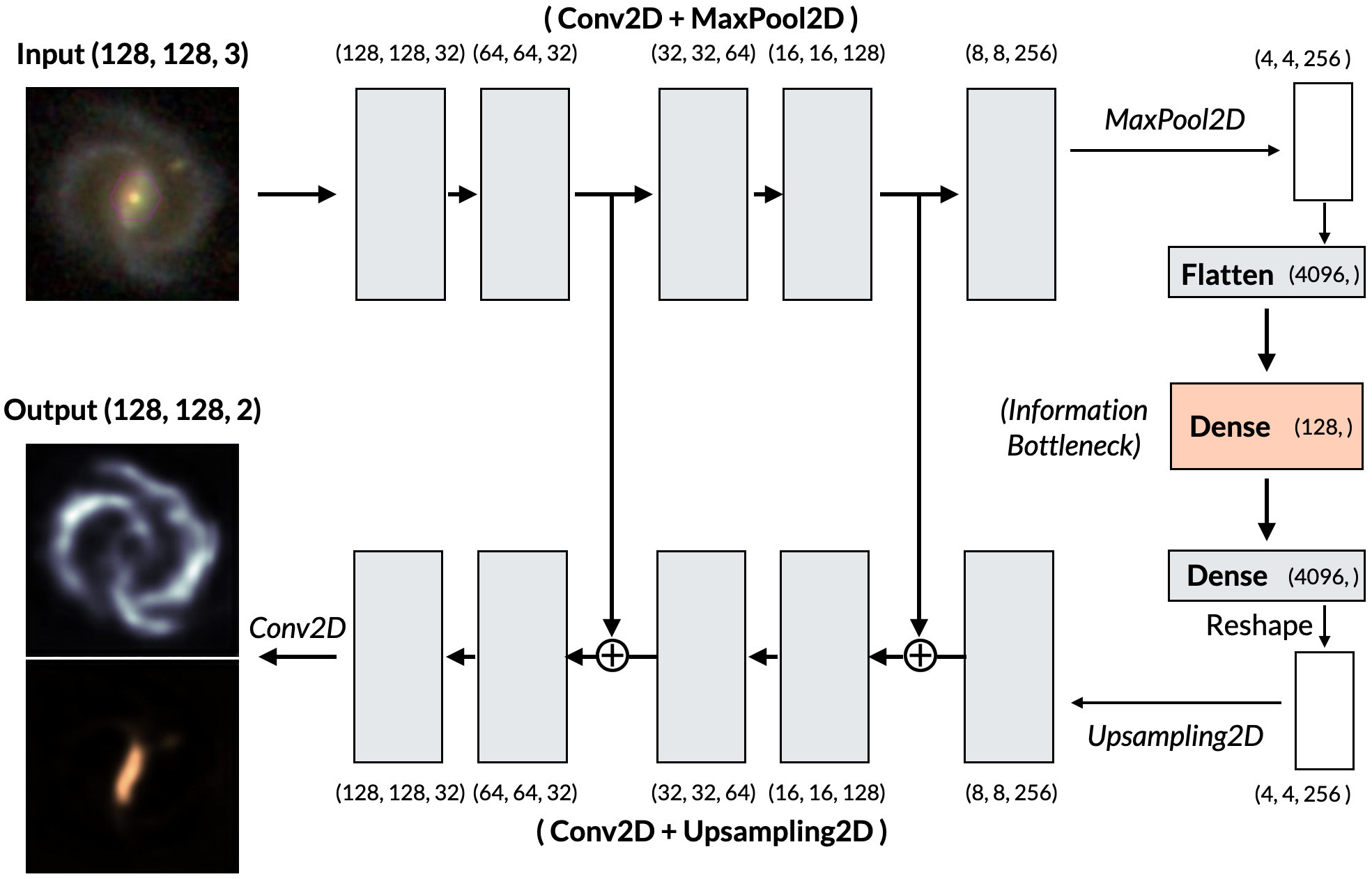}
    \caption{A simplified schematic of our U-Net architecture, showing all core layers, with arrows indicating the direction of data progression. Tuples in parentheses indicate the output dimensions of the indicated layer, with the rightmost digit indicating the number of channels. Each of the five top-most blocks consist of alternating Conv2D and MaxPool2D layers; each of the five bottom-most blocks consist of alternating Conv2D and Upsampling2D layers.}
    \label{fig:unetdiagram}
\end{figure*}

\subsubsection{Model Architecture}

A U-Net is a type of specialised convolutional neural network specifically designed for image segmentation. A full treatment of CNNs is beyond the scope of this paper, however we will give a brief conceptual overview of their operation below (for a thorough treatment of CNNs, see \citealt{lecun2015,goodfellow2016,chollet2021}). In a typical CNN, for instance one designed for image classification, the model takes an image as an input, then applies successive layers of convolutional filters to progressively extract abstract features from the images, as well as pooling layers to progressively downsample the outputs. In the case of image classification, the final output is often an array of probabilities; one for each possible image category. The values for these output probabilities are also known as confidences.

In the case of the U-Net, the input and output are both images. U-Nets can be thought of as a regular CNN turned back on itself. The input image is first progressively convolved and downsampled into a latent information bottleneck, before a new image is reconstructed through the use of additional convolutions and upsampling layers. Figure~\ref{fig:unetdiagram} shows a simplified schematic of our core U-Net archiecture, clearly showing how the two halves downsample the input image, then reconstruct the output images. At first glance, this architecture appears to resemble the structure of a classical autoencoder (indeed, the principle of learning a low-dimensional representation is exactly the same). However, what distinguishes U-Nets from an autoencoders the presence of residual links, also known as skip connections. These links preserve a copy of the outputs of a given layer, such that they can be reincorporated into the input of a downstream layer. Not only do these residual layers allow for the training of much deeper networks through improved gradient propagation \citep{he2015} but, in the contexts of U-Nets, they also help to condition the upsampling process by directly retaining the high-level features learned in the downsampling process. This is a crucial reason behind the tremendous success of U-Nets at image segmentation \citep{minaee2020}.

Our core architecture, shared by both the RGB colour and monochromatic U-Nets, consists of eleven convolutional layers (five each for the downsampling and upsampling process, plus one for the output layer). Each convolutional layer employs a kernel size of 3. \texttt{LeakyReLU} activation functions are used throughout, with sigmoid activation for the output layer. Batch normalisation is also used for all convolutional layers except for the output layer. There are two residual links connecting convolutional layers with equal output dimensions via element-wise addition (this is done using \textsc{Keras}' \texttt{Add} layer). The information bottleneck is a fully-connected \texttt{Dense} layer with 128 nodes; this is the hence dimension of the smallest latent space of the model. The only architectural difference between our two U-Nets is the number of channels in the input layer, which results in an extremely negligible increase in the number of trainable parameters for the colour U-Net (2,459,586), compared to the single-channel input, monochrome U-Net (2,459,522).

\subsubsection{Data Preprocessing \& Model Training}

\begin{figure*}
    \includegraphics[scale=0.87]{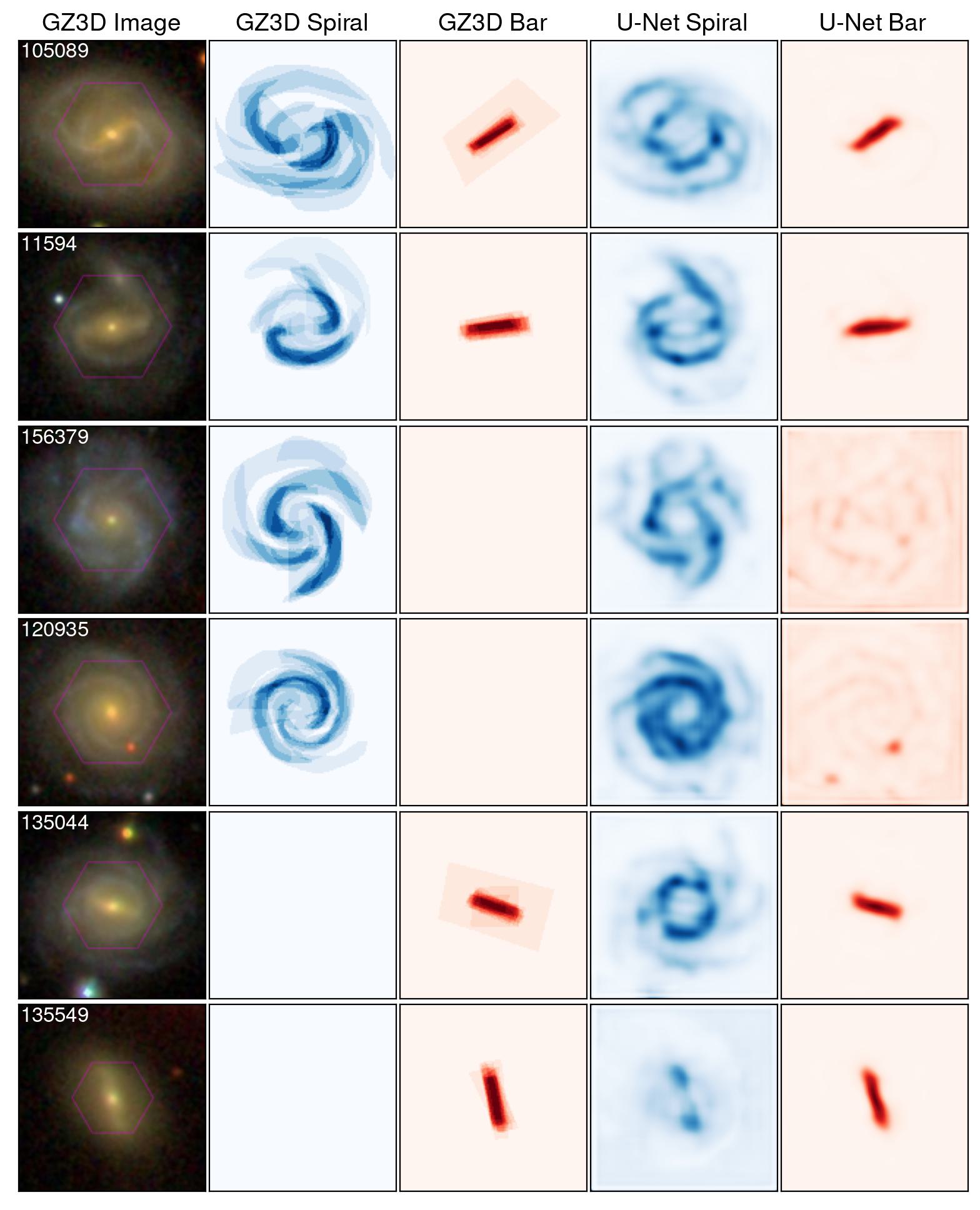}
    \caption{Example application of the U-Net to six randomly selected galaxies from the GZ3D test set; two barred spirals, two unbarred spirals, and two barred galaxies without GZ3D spiral masks. Each row shows the input GZ3D colour cutout, the volunteer-drawn GZ3D spiral and bar masks, and the subsequent predicted spiral and bar masks as directly outputted by the U-Net. The image cutouts are annotated with their GZ3D ID in the top-left. Different shades in the GZ3D count masks correspond to different numbers of classifiers; darker colours indicate higher degrees of overlap. The U-Net outputs attempt to emulate the count masks as closely as possible, but are much smoother since the pixel values are continuous.}
    \label{fig:gz3dexample}
\end{figure*}

\begin{figure*}
    \includegraphics[scale=0.41]{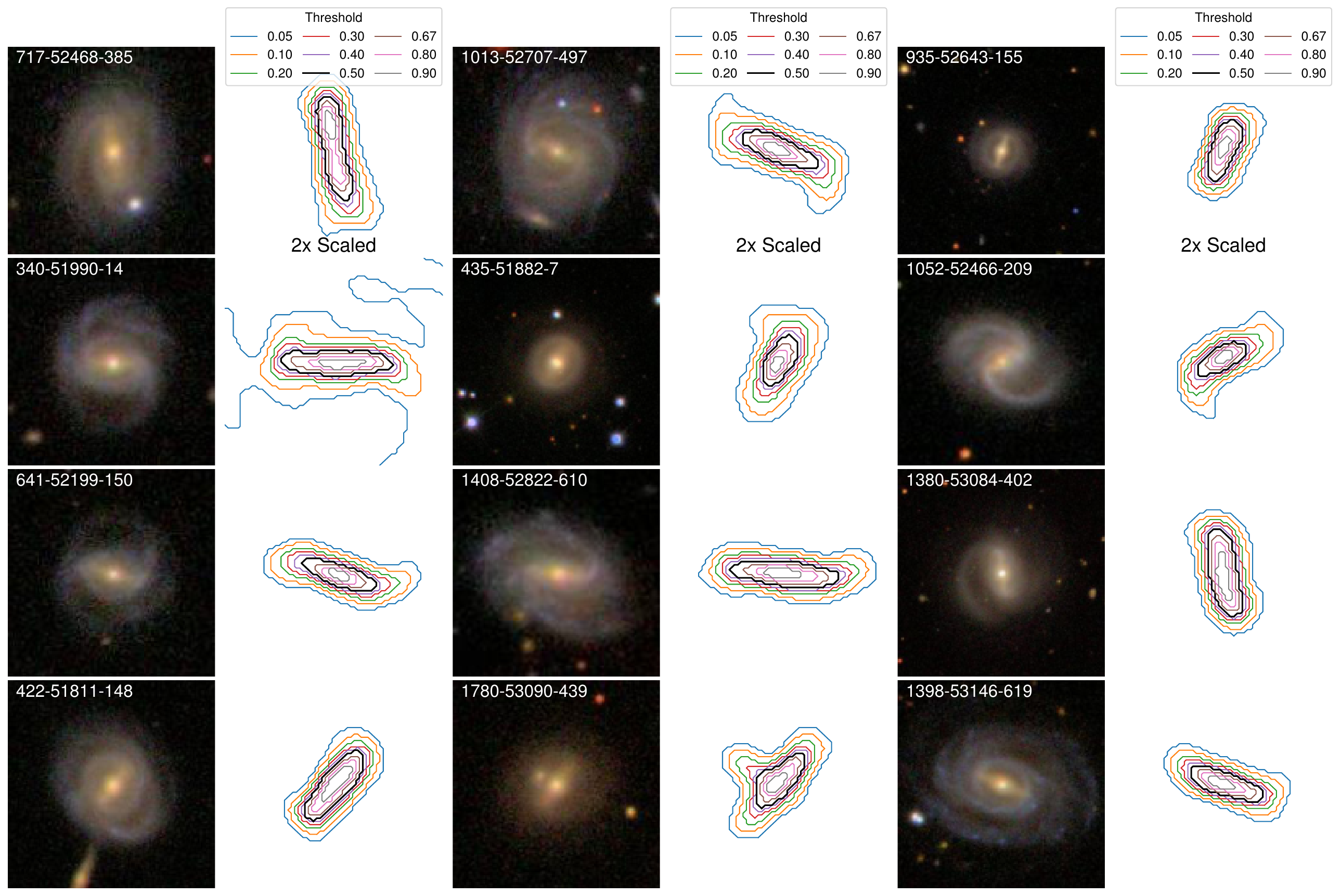}
    \caption{Contours defined by applying different local thresholds to the predicted bar masks for a random selection of GZ3D cutouts from the test set. Note that the image scale for the contours has been magnified by a factor of two.}
    \label{fig:barcontours}
\end{figure*}

The GZ3D masks were obtained through crowdsourcing as part of the wider Zooniverse citizen-science project. As previously described in Section~\ref{ss:gz3d}, each contributor was tasked with drawing pixel masks corresponding to the locations of bars and spirals. The final mask is hence determined through aggregating all the masks together (see \citealt{masters2021} for further details). As such, there are inevitably regions where the masks from multiple individual classifiers overlap. The degree of overlap can be considered as a proxy of confidence, in the sense that multiple volunteer contributors are in agreement. This information therefore ought to be incorporated into the deep learning model. Indeed, a key difference of our U-Nets is that instead of directly outputting a discrete-valued integer pixel mask, as is commonly done in image segmentation, our U-Nets output continuous values for each pixel, such that larger values indicate higher degrees of confidence. This is akin to pixel-level regression rather than pixel-level classification. This method allows us to account for, and ultimately preserve, the differing degrees of overlap in the aggregate count masks in the GZ3D. Were the U-Net instead configured to output integer values, then this information would be lost. Another benefit of configuring the U-Nets in this manner is that this allows arbitrary thresholds to be applied to the outputs of the model. If the model were instead trained with binary pixel masks, then a threshold must have been pre-applied to the training data. All subsequent applications of the model would then be contingent on this pre-applied threshold. Instead, our approach is for the U-Net to reproduce the GZ3D count masks.

The data preprocessing first commences by processing the raw fits files for each of the 7,965 galaxies in our GZ3D sub-sample, publicly accessible as part of the SDSS DR17 catalogue \citep{abdurrouf2022}. The RGB colour cutouts of each GZ3D galaxy were generated from \textit{gri} SDSS imaging with a pixel scale of 0.099 arcsec/pixel at a size of 525x525 pixels, corresponding to an angular field of view of 52 arcsecs \citep{masters2021}. Note that we will hereafter refer to these colour images as RGB, keeping in mind that they correspond to \textit{gri} photometric bands. The count masks share the same initial size of 525x525 pixels. We first apply a centre crop with a border width of 32 pixels, resulting in images of size 461x461 pixels; this serves to trim part of the background and better focus in on the galaxy, while subsequently enlarging the masks. The images are then resized to our U-Net's input size of 128x128 pixels using the default bi-cubic interpolation settings with Python's \textsc{Pillow} package. All pixel values, including for the count masks, are rescaled such that they are within the range 0 to 1, as is conventional for deep learning. In the case of the images, all pixel values are divided by 255. For the count masks, each pixel value is instead divided by the largest pixel value out of all masks, which by definition is the maximum number of volunteer classifiers $N_{\text{max}}$. Note that the bar and spiral count masks are rescaled separately. Furthermore, under this formulation, the count mask values are not probabilities, but simply rescaled aggregate counts. For instance, if for one galaxy 4 out of 4 classifiers marked a given pixel as ''barred'', then it will have the same rescaled pixel value 4/$N_{\text{max}}$ as one in another galaxy that 4 out of 8 classifiers marked as barred. An alternative formulation is to consider count fractions, in which the previous 4/4 and 4/8 examples would therefore have pixel values of 1 and 0.5 respectively. However, this would give a disproportionately high weight to masks with low volunteer counts. Thus, in order to best replicate the count masks, we just perform linear scaling.

We partitioned the images and labels into train and test sets using a ratio of 85:15; that is, 85\% of the 7,965 galaxies are used to train the model, and the remaining 15\% are used to evaluate the model. These same partitions were used to train the colour and monochromatic U-Net. In the case of the monochromatic U-Net, the galaxy images were converted to monochrome using \texttt{Pillow}. The models are trained for up to 100 epochs with \textsc{Keras}'s \texttt{EarlyStopping} callback. This callback terminates the training if the loss has not improved over the previous $n$ epochs, where $n$ is commonly referred to as the patience. We use a patience of 11.

\begin{figure*}
    \includegraphics[scale=0.83]{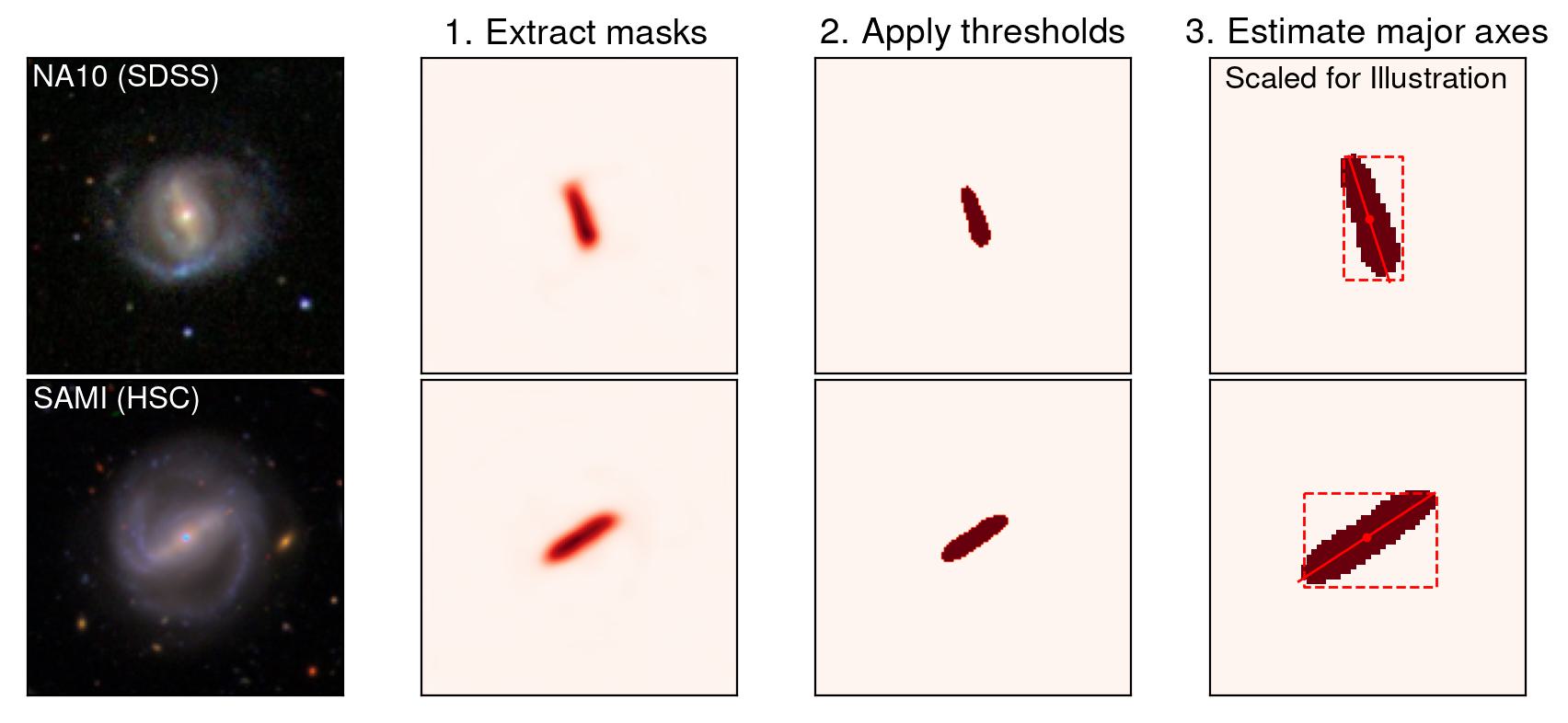}
    \caption{Overview of the full sequence of steps in our bar length estimation pipeline. First, the U-Net is applied to an image of a galaxy. The U-Net subsequently outputs a smooth bar mask with continuous pixel values that are proxies for different levels of confidence. The next step is to choose a threshold with which to convert this continuous bar mask into a binary pixel mask. The final step involves ellipse fitting of the integer-valued pixel mask to calculate the length of the major axis.}
    \label{fig:barlengthalg}
\end{figure*}

Figure~\ref{fig:gz3dexample} demonstrates the capabilities of the fully-trained U-Net at replicating the GZ3D count masks. One immediately noticeable difference is the smoothness of the predicted spiral masks, which reflect the interpolation utilised in the upsampling process. It is also clear that the predicted spiral masks are highlighting regions of the image that correspond to the presence of a spiral arm rather than tracing out the spiral arm, hence the more patchy appearance. This is most prominently seen in the fifth row of Figure~\ref{fig:gz3dexample}, in which the predicted spiral mask picks out the faint spiral arms despite there not being any provided mask in the GZ3D. Even though this is a desirable outcome, issues with the GZ3D masks (or lack thereof) like this do adversely impact the training of the model. We will discuss these errors and the performance of the model on corrupt or missing masks in Appendix~\ref{app:inputs}. Figure~\ref{fig:gz3dexample} also illustrates the importance of having configured the U-Net to output separate masks for bars and spirals. In general, the predicted bar and spiral masks are well separated. In some cases the spiral mask clearly leave a gap for where the bar should be. The predicted bar masks are smooth and correctly match the orientation of the bars, completely eliminating the blocky nature of the GZ3D bar masks while retaining differences in confidence levels. In the third and fourth rows of Figure~\ref{fig:gz3dexample}, which show the outputs for unbarred galaxies, the bar masks have substantially smaller pixel values, instead tracing an echo of the galaxy. Interestingly, we see in the fourth example that the bar mask appears to highlight the locations of the two foreground stars. This suggests that the bar mask outputs may be sensitive to redder colours, as bars are generally redder than the rest of the host galaxy, let alone predominantly blue, star-forming spiral arms. We will discuss these points, and how the monochromatic U-Net fares in the absence of colour, in Section~\ref{ss:rgbmono}.

\subsection{Bar Length Estimation}

As aforementioned, the direct outputs of the U-Net are not discrete, binary integer masks, but continuous values that correspond to degrees of confidence, as has been illustrated in Figure~\ref{fig:gz3dexample}. Indeed, since the U-Net was trained directly on the GZ3D count masks, the total range of pixel values within a given mask will vary from galaxy to galaxy. This includes cases where the outputted pixel values are extremely tiny, notably in the two unbarred galaxies in the middle two rows Figure~\ref{fig:gz3dexample}. It is therefore desirable to discard masks whose pixel values are too small. Our bar estimation pipeline thus includes two thresholds:
\begin{itemize}
\item A \textit{global threshold} in order to discard bar masks whose pixel values are too small. 
\item And a \textit{local threshold} to decide which pixels to keep in the current bar mask.
\end{itemize}
Specifically, a bar mask is chosen if its largest pixel is at least some fraction of the maximum pixel value across all bar masks for the given dataset. This is the role of the global threshold, the purpose being to enforce a minimum threshold for the quality of a predicted bar mask (c.f. middle two rows of Figure~\ref{fig:gz3dexample}. For this study we set it at 10\%. Once we've selected a mask, we need to convert it from the continuous-valued mask into a discrete, 0-1 binary mask. This is the role of the local threshold, which is defined as some fraction of the maximum pixel value in the mask. Pixels greater than this value are set to 1, the rest are set to 0. Here, a threshold of 50\% means that only pixels whose value is greater than half the value of the largest pixel (in the current mask) are set to 1, with all other pixels set to 0. Given that the U-Net is trained to reproduce the count masks, this is analogous to choosing to retain only the regions of overlap shaded by more than half of the volunteer classifiers (for a given galaxy). For this study, we set the local threshold at 50\%. Defining both these local and global thresholds as fractions allows our algorithmic approach to automatically process bar masks for any arbitrary set of galaxy images, irrespective of the ranges of pixel values.

Figure~\ref{fig:barcontours} shows the impacts of applying different local thresholds to the predicted bar masks. In general, higher thresholds extract progressively smaller regions, while lower thresholds are more susceptible to error and variation, in some cases tracing the spiral arm. The bottom-most example of the central column in Figure~\ref{fig:barcontours} shows that the U-Net is susceptible to companion sources, but a suitably high threshold can mitigate this. However, at very high thresholds, it can be seen that the extent of the bar is severely underestimated. This motivates our use of 0.5 as a local threshold. It should be stressed that the local threshold is merely a way to define consistent quantiles of the U-Net outputs. This is since the count masks have different ranges of values as not all galaxies were classified by the same number of volunteer classifiers. If instead the count masks were converted to probabilities, then a local threshold would still be necessary, albeit instead defined as a fixed probability value (e.g. 0.5) instead of a percentage.

\begin{figure*}
    \includegraphics[scale=0.735]{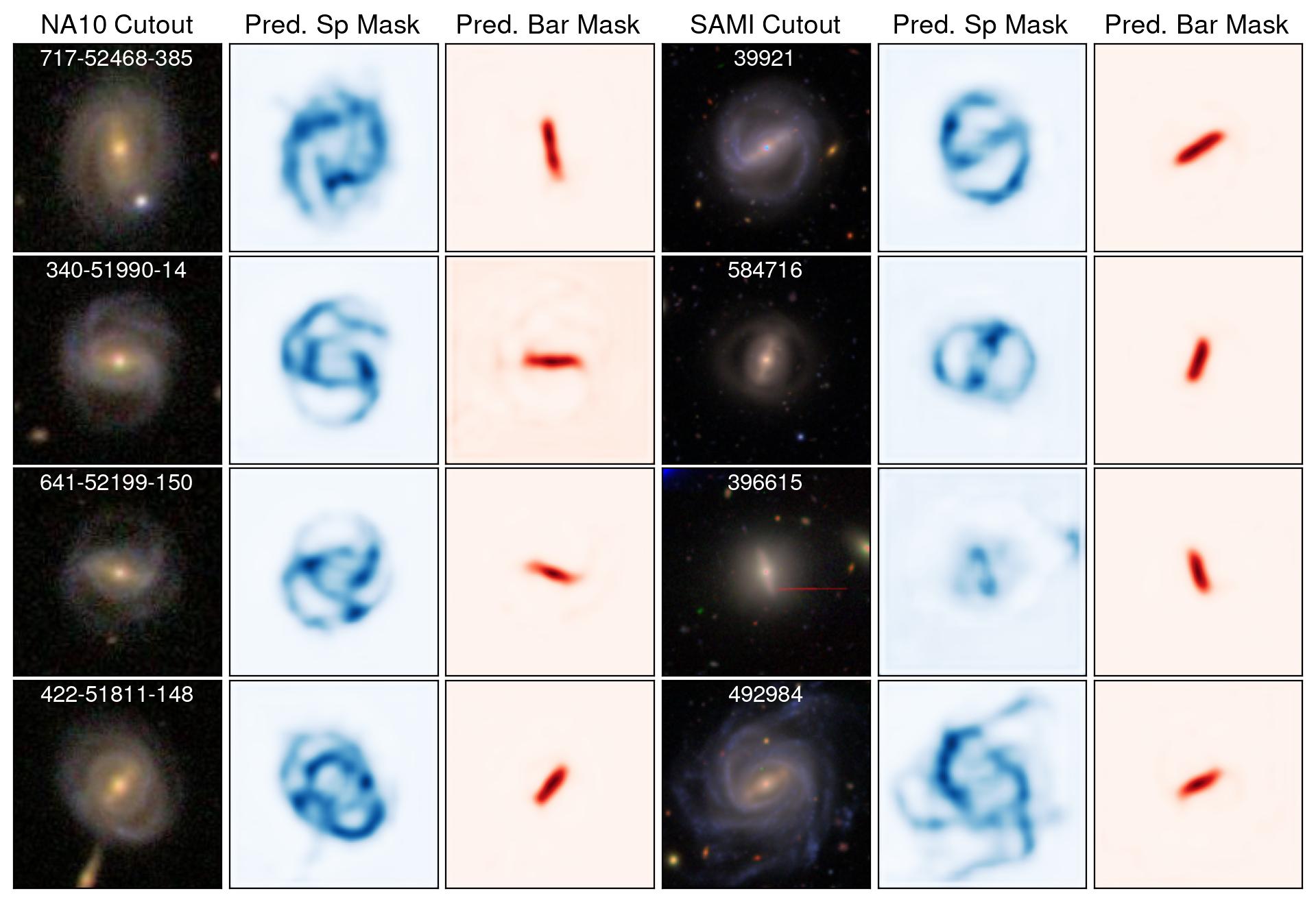}
    \caption{Predicted bar and spiral masks as directly outputted by the U-Net for a random selection of NA10 galaxies and SAMI galaxies. NA10 and SAMI cutouts are annotated with their spID and SAMI catalogue IDs respectively.}
    \label{fig:examplemasks}
\end{figure*}

Once the binary mask is obtained via the local threshold, the next step is to perform ellipse fitting. For this, we utilise \texttt{scikit-image}'s \texttt{measure} module to automatically perform the ellipse fitting, and measure the length of the bar by measuring the major axes of the fitted ellipse \citep{vanderwalt2014}. More specifically, we use the \texttt{label} function to detect and label connected pixel regions (i.e. neighbouring pixels that share the same value) in the resulting binary pixel bar mask. This is followed by the \texttt{regionprops} function, which enables us to subsequently estimate the bar length based on the length of the major axis of the fitted ellipse assigned to this connected region. Our choice of using the routines from the \texttt{scikit-image} package to carry out the ellipse fitting is motivated by the fact that these functions are specifically designed for discrete integer-valued masks. We note that this choice is by no means exclusive, and that the ellipse fitting of the pixel mask can feasibly be conducted with other image processing libraries such as \textsc{opencv}. By this stage in the overall estimation pipeline, the estimated bar length is in units of pixels, which can be converted into a physical length based on the (per-pixel) physical scale of the corresponding image. Figure~\ref{fig:barlengthalg} illustrates the full sequence of steps for two example barred galaxies from the NA10 and SAMI datasets. The entire pipeline is fully automated, requiring only the global and local thresholds for bar mask filtering and extraction respectively, which in our case were set to 0.1 and 0.5.

It is worth noting that our choice of carrying out binary-pixel mask extraction by applying a threshold is merely one approach to estimating the bar length. It is also feasible to obtain the pixel mask through using traditional isophotal ellipise fitting of the smooth masks directly outputted by the U-Net (i.e. after Step 1 in Fig~\ref{fig:barlengthalg}), or indeed through applying ellipse fitting to the original galaxy image in the first place. That said, compared to isophotal ellipse fitting – and parametric fitting methods in general – our U-Net approach to bar detection has several key benefits and drawbacks. One of the immediate benefits is that it is computationally efficient; the bar mask is directly obtained without the need for an iterative fitting process. Furthermore, there is no need to specify initial conditions as there are no free parameters (c.f. \citealt{aguerri2009}). This allows us to apply the U-Net to readily and rapidly extract bar masks for thousands of galaxies across different datasets in less than a minute; specifically, $\approx$10 seconds for the NA10 galaxies, and $\approx$16 seconds for the SAMI galaxies. Our use of \texttt{scikit-image}'s \texttt{measure} model enables us to quickly determine many physical metrics for the bar apart from its semi-major axis, such as spatial moments and pixel area, the latter of which is extremely useful for determining mass fractions (we will discuss additional applications of the U-Net in Section~\ref{ss:modeldisc}). That the U-Net is trained solely on the images of galaxies (paired with their respective spiral and bar masks) without any additional hard-coded assumptions or auxiliary information is a fundamental advantage of deep learning methods in general. However, due to the sole reliance on the training data, this can lead to limitations in the model's ability to generalise. This is best illustrated in the previously discussed example of Figure~\ref{fig:gz3dexample}, where the bar mask is sensitive to red foreground stars, or in Figure~\ref{fig:barcontours} where, at low thresholds, the predicted ``bar'' mask appears to protrude along spiral arms. The concept of a ``bar'' is, from the point of view of the U-Net, simply the features of a galaxy image that correspond to regions that have been labelled as such across the thousands of GZ3D bar masks it has been trained on. As such there is the potential for greater uncertainty when compared to a parametric approach, such as isophotal ellipse fitting or Fourier analysis.

\section{Results}
\label{s:results}
    
\begin{figure*}
    \includegraphics[scale=0.86]{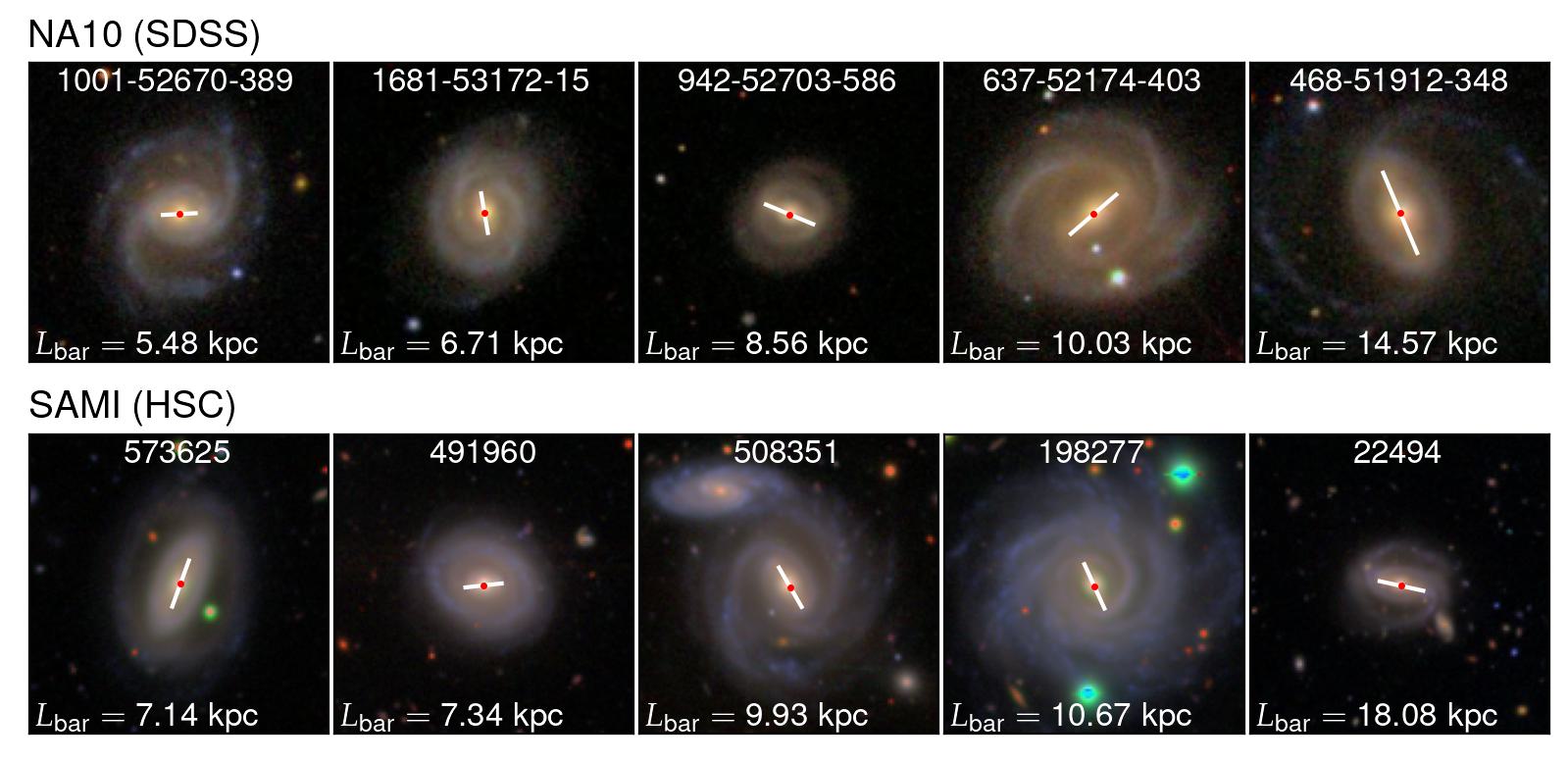}
    \caption{Central coordinates (denoted by red markers) and estimated bar lengths $L_{\text{bar}}$ (shown with white lines oriented according to the predicted bar mask) for a random selection of NA10 and SAMI galaxies is ascending order. Cutouts annotated as in Figure~\ref{fig:examplemasks}.}
    \label{fig:examplelengths}
\end{figure*}

\begin{figure*}
    \includegraphics[scale=0.735]{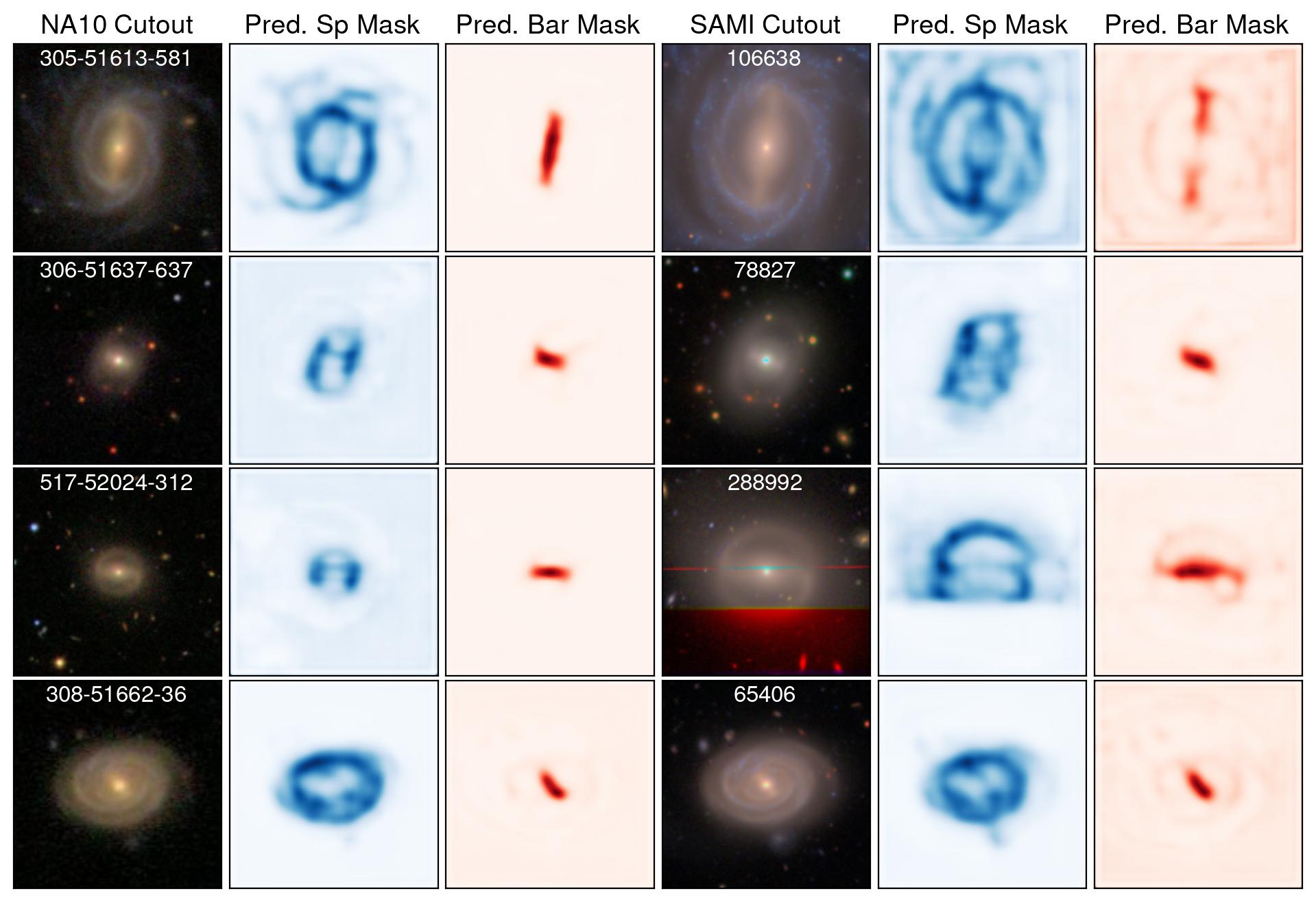}
    \caption{Predicted bar and spiral masks as directly outputted by the U-Net for a random selection of NA10 galaxies and SAMI galaxies. NA10 and SAMI cutouts are annotated with their spID and SAMI catalogue IDs respectively.}
    \label{fig:examplemasks-samegal}
\end{figure*}

\subsection{Application of the U-Net}

We first apply our fully-trained U-Net to extract bar masks for galaxies from the NA10 morphological catalogue and SAMI DR3. As an additional measure of confidence, before processing each image with the U-Net, we verified that the image is barred by classifying it with our bar CNN. This extra step did not affect the SAMI bars (which were classified by the model in question), but this did exclude approximately 2.5\% of the known NA10 bars, which is well within the 83\% accuracy of the bar model (see \citealt{cavanagh2021}).

Figure~\ref{fig:examplemasks} illustrates the predicted spiral and bar masks for a selection of galaxies from NA10 and SAMI. It can be seen that the model performs consistently well on the two different source imagery. Despite being trained on SDSS imaging, the U-Net is also able to extract bar masks for HSC imaging. There is, however, a higher degree of overlap in the predicted spiral masks with the central bar region. The predicted spiral mask is also sensitive to galaxy rings, as shown with the predicted masks for the SAMI galaxy with ID 584716. This can be considered an undesirable side effect given the model is ostensibly trained to solely extract spiral arms, but it nevertheless demonstrates how the U-Net works in principle by responding to features in the image. That said, spiral arms are an incredibly diverse morphological feature and can manifest in a variety of structures and forms, whilst bars are comparatively more uniform. Indeed, the predicted bar masks in Figure~\ref{fig:examplemasks} are all smooth and correctly oriented. It is also encouraging that the extracted bar masks all highlight a single region in the image, despite some instances of visual interference or neighbouring sources, such as in the NA10 galaxy 422-51811-148.

\begin{figure*}
    \centering
    \includegraphics[scale=0.53]{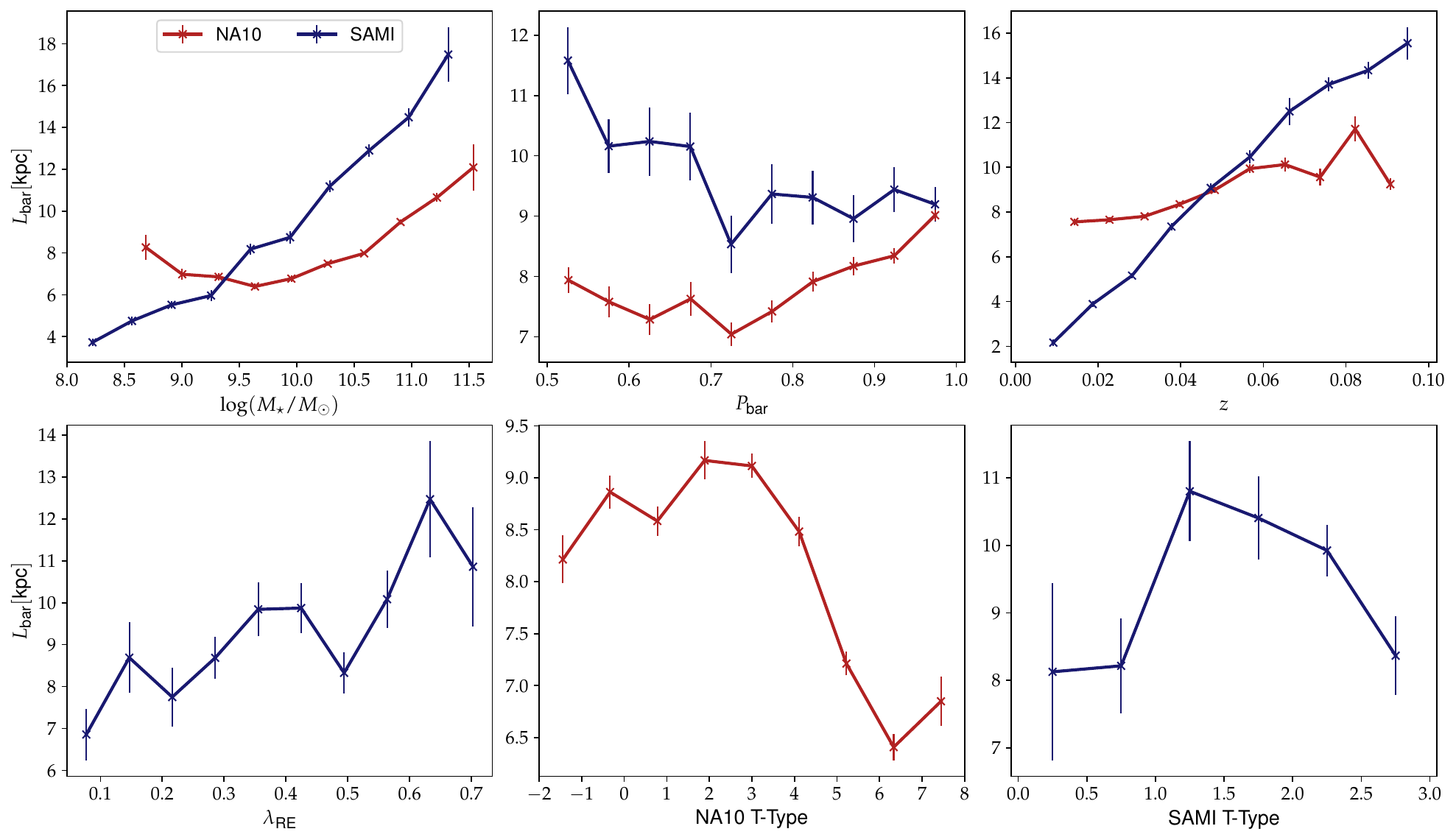}
    \caption{The absolute bar length $L_{\text{bar}}$ in terms of stellar mass, bar classification confidence $P_{\text{bar}}$, redshift $z$, the spin parameter proxy $\lambda_{R_e}$, NA10 T-Types and SAMI T-Types. NA10 and SAMI galaxies are coloured in dark red and dark blue respectively. Error bars denote one standard error $\sigma$.}
    \label{fig:aio-lb}
\end{figure*}

Having now obtained the bar masks for the NA10 and SAMI galaxies, the bar length estimation pipeline can proceed. Figure~\ref{fig:examplelengths} shows a selection of galaxies marked with the estimated bar length based on the fitted ellipse as per \texttt{scikit-image}'s \texttt{measure} module. The SDSS imaging all have the same physical scale of 50 kpc, while the HSC cutouts vary in physical scale. In general, the markers are well aligned with the galaxy centres, and the predicted semi-major axes are correctly oriented. There are some instances where the centre is slightly offset. This is possibly due to one side of the predicted bar mask having a higher overall confidence than the other side, subsequently offsetting fitted ellipse. This does not appear to significantly affect the overall length of the bar. It is also worth keeping in mind the inherent uncertainty of where the bar ends and where the spiral arm begins (as previously shown in Figure~\ref{fig:barcontours}), which could lead to an overestimated $L_{\text{bar}}$ in the case of strong spiral galaxies.

Figure~\ref{fig:examplemasks} shows that the U-Net can successfully process both SDSS and HSC imaging. However, there remains the question of consistency, and whether there are any major differences. We can examine this by applying the U-Net to extract spirals and bars from SDSS and HSC imaging of the same galaxy. There are exactly four barred galaxies in the NA10 catalogue that are also present in our SAMI DR3 subsample. These are shown in Figure~\ref{fig:examplemasks-samegal}, along with the predicted spiral and bar masks. The first immediate difference is the different angular resolutions, however the U-Net is nevertheless able to extract masks, even at the edges of the image. However, the bar masks are less uniform. This is especially problematic in the case of galaxy 106638, where the U-Net model has difficulty extracting the centre of the bar, possibly due to it being obfuscated by the large central bulge. In 288992 and 65406, the predicted bar masks appear to trace the spiral arms (this effect can be removed with a suitably large local threshold). It is interesting to note how the U-Net handles the image artefacts in 288992; the horizontal line through the centre of the galaxy does not appear to affect the predicted bar mask, while the model does not output anything in the red band artefact region. The estimated bar lengths from the predicted bar masks are 15.6, 6.83, 8.19 and 7.61 kpc for the SDSS imaging, and 0 (skipped due to being below the global threshold), 5.78, 6.33 and 8.38 kpc for the HSC imaging respectively. The discrepancies in bar length directly follow from the discrepancies in the output bar masks and, more specifically, the threshold contours.

\subsection{Physical Properties}

Bar length, also denoted $L_{\text{bar}}$, is a key physical property of stellar bars. Understanding how bar length varies with respect to the properties of the host galaxy can yield important insights into the processes that govern the growth, evolution and impacts of bars \citep{combes1993, gadotti2006, hoyle2011, guo2019, fraser-mckelvie2020barlength}. The bar length $L_{\text{bar}}$ is one of several phenomenological criteria often used to judge the ``strength'' of a stellar bar \citep{aguerri1998, guo2019, geron2021}. Previous studies have shown that the bar length tends to scale with the mass of the host galaxy, and may also change depending on the morphology of the galaxy with larger bars in early-types and shorter bars in late-types \citep{erwin2005, diaz-garcia2016, erwin2019}. It is worth noting that the absolute bar length $L_{\text{bar}}$ is not necessarily the ideal quantity to meaningfully study the lengths of bars across different galaxies with different physical properties. In particular, a large bar in a small galaxy may share the same, absolute length as a small bar in a large galaxy, but they are clearly sized differently in relation to their host galaxy. To take the size of the host galaxy into consideration, we can define a normalised, relative bar length by considering the dimensionless quantity $L_{\text{bar}} / R_d$ where $R_d$ is the scale radius of the galaxy disk, also known as the scale length. In the case of the SAMI galaxies, $R_d$ is calculated from the effective radius $R_e$ as $R_d \approx R_e / 1.68$. For the NA10 galaxies, $R_d$ is similarly calculated from the effective radius $R_e$, which is approximated from the Petrosian half-flux radius $R_{50}$ and concentration $R_{90}/R_{50}$ using the method in \citet{graham2005}.

\begin{figure*}
    \centering
    \includegraphics[scale=0.53]{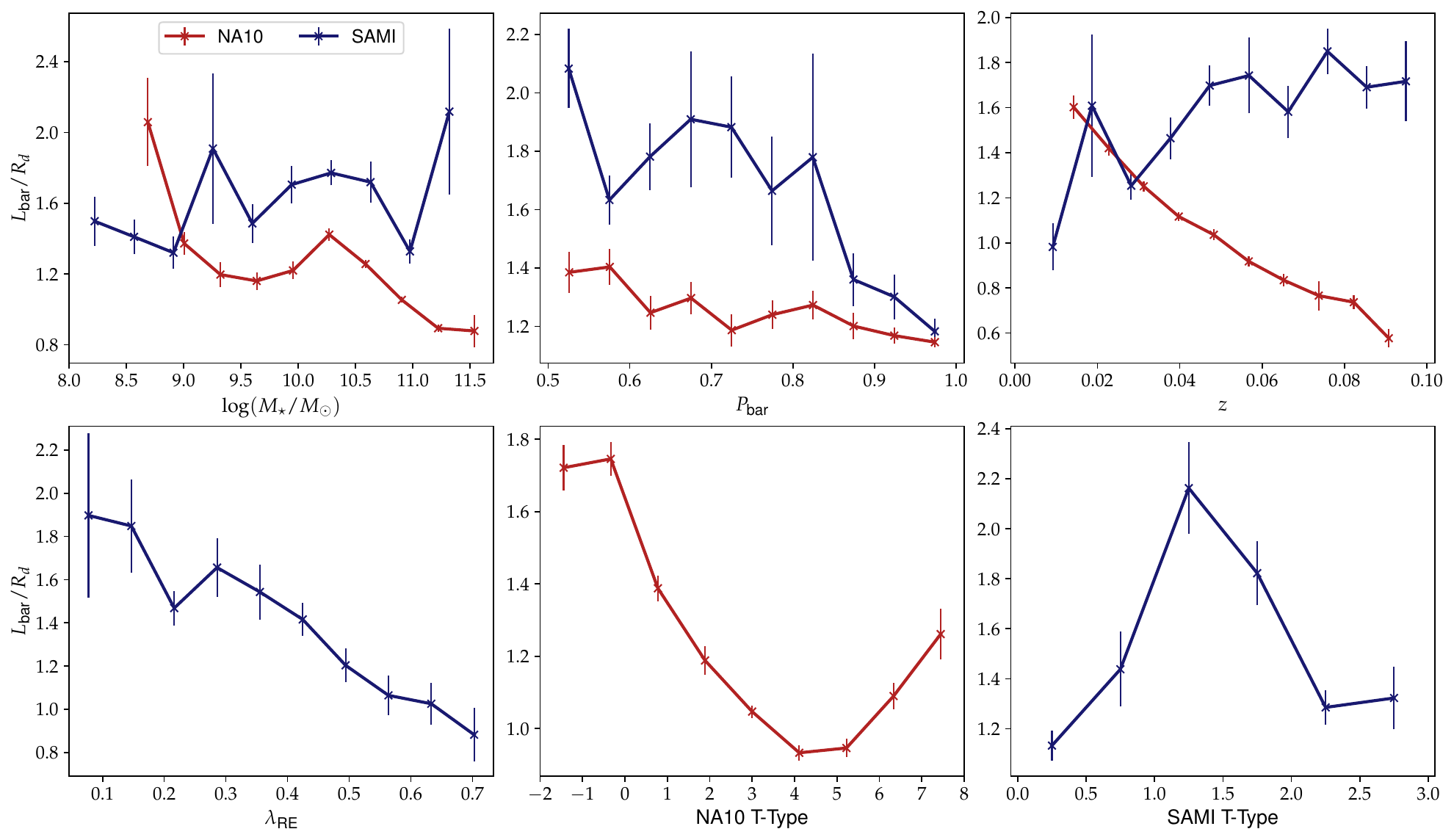}
    \caption{The same plots as in Figure~\ref{fig:aio-lb}, but instead showing the normalised, relative bar length $L_{\text{bar}} / R_d$.}
    \label{fig:aio-lbrd}
\end{figure*}

Figure~\ref{fig:aio-lb} shows a series of plots for the absolute bar length $L_{\text{bar}}$ with respect to stellar mass, the bar classification confidence $P_{\text{bar}}$ based on the bar CNN model from \citet{cavanagh2022}, redshift, $\lambda_{R_e}$, as well as morphology in terms of the NA10 T-Types (themselves based on Hubble T-Types), as well as the SAMI T-Types described in \citet{cortese2016}. Likewise, Figure~\ref{fig:aio-lbrd} examines the same properties for the relative bar length $L_{\text{bar}}/R_d$. It can be seen that $L_{\text{bar}}$ increases strongly with stellar mass $\log(M_\star/M_\odot)$, a trend that has been well established in observations \citep{kormendy1979, erwin2005, diaz-garcia2016}. However, when considering the size of the host galaxy, the trends are mixed between the two datasets. In the case of NA10, there is a significant drop in the relative bar length $L_{\text{bar}}/R_d$ for high mass galaxies; bars are longer in high mass galaxies, but shorter with respect to the host galaxy. On the other hand, the relative bar length trends upwards for the SAMI galaxies, with significant variation. That said, it is important to note that the two datasets sample different mass ranges, and that the mass distributions are different. Studies have shown that disc scale radius increases with stellar mass, suggesting that as a host galaxy grows, so too does its bar, albeit not necessarily with respect to its disc \citep{diaz-garcia2016,kruk2018,fraser-mckelvie2020barlength,rosas-guevara2022}. This could also be reflective of higher bulge-to-disk ratios for more massive galaxies. It is argued in \citet{erwin2019} that difficulties in detecting shorter bars may result in their underestimation in low mass galaxies. \citet{erwin2018} also makes the argument that limited image resolution affects the ability to detect bars. This could be a reason as to why $L_{\text{bar}}/R_d$ remains high for low mass galaxies in Figure~\ref{fig:aio-lbrd}, in the sense that the U-Net is better suited to extracting the largest of bars while missing smaller bars. We will return to this point when discussing galaxy size in more detail in Section~\ref{s:discussion}.

Comparing $L_{\text{bar}}$ with the classification confidence $P_{\text{bar}}$ provides a means of assessing biases. Here, $P_{\text{bar}}$ is the confidence that the model is barred, as predicted by the bar CNN (this is separate from the U-Net). In the case of NA10, there is a slight increase in $L_{\text{bar}}$ with high $P_{\text{bar}}$ galaxies, suggesting that longer bars are more confidently classified. For SAMI this trend is reversed, with longer bars in low $P_{\text{bar}}$ galaxies. This is likely reflective of the different source imaging and, subsequently, different uncertainties, similar to what was observed in Figure~\ref{fig:barcontours} with lower thresholds. The predicted bar mask for a low-confidence bar galaxy may well be larger than for a high-confidence bar galaxy, not because the bar is longer, but because the extracted bar mask is itself larger out due to the greater uncertainty. Indeed, Figure~\ref{fig:aio-lbrd} shows that $L_{\text{bar}}/R_d$ is highest for low $P_{\text{bar}}$ galaxies in both the NA10 and SAMI datasets. This is, of course, subject to the interpretation of $P_{\text{bar}}$ as an indicator for how easy it is to identify a galaxy as barred (see also \citet{cavanagh2022} where it is argued that $P_{\text{bar}}$ is a proxy for bar strength).

Figures~\ref{fig:aio-lb}~and~\ref{fig:aio-lbrd} also show how the bar length varies terms of redshift $z$ and the spin parameter proxy $\lambda_{R_e}$. In the case of NA10, we can observe that while $L_{\text{bar}}$ increases with increasing $z$, $L_{\text{bar}}/R_d$ dramatically decreases. This is in stark contrast to SAMI, where we see an increase in both $L_{\text{bar}}$ and $L_{\text{bar}}/R_d$ with $z$. There are many factors that can influence the lengths of bars with redshift, not least the morphological evolution of the host galaxy. However, previous studies examining bars out to high redshifts have found that bar lengths (both absolute and normalised length) do not show any significant changes with redshift \citep{sheth2008,kim2021}, implying that bars scale proportionally with their host galaxy as both evolve over time. The results of Figures~\ref{fig:aio-lb}~and~\ref{fig:aio-lbrd} suggest that this might not be so clear-cut.

\begin{figure}
    \centering
    \includegraphics[scale=0.58]{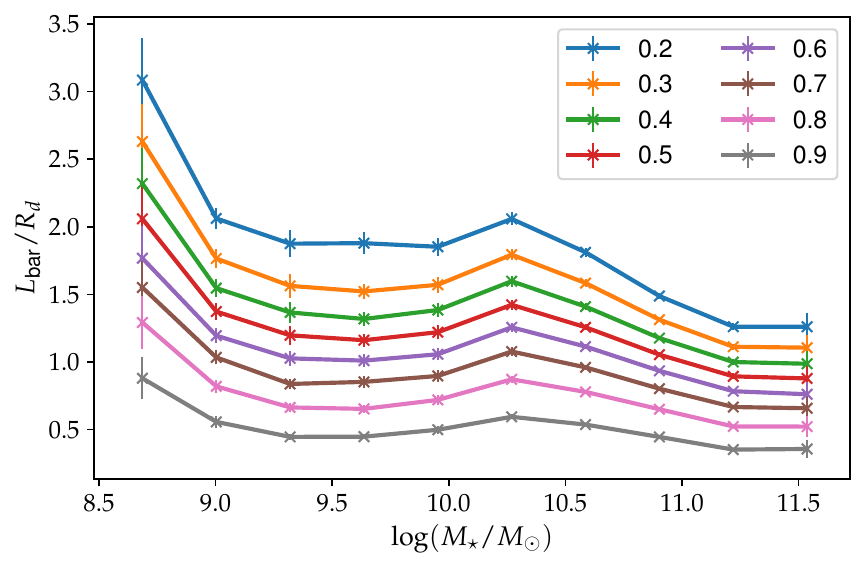}
    \caption{The dependence of the relative bar length $L_{\text{bar}}/R_d$ on stellar mass $\log(M_\star/M_\odot)$ for different local thresholds as applied to the predicted bar mask.}
    \label{fig:lbthres}
\end{figure}

The spin parameter proxy $\lambda_{R_e}$, originally introduced in \citet{emsellem2007}, is a key kinematic property of galaxies. This dimensionless parameter acts as a proxy for the baryon projected specific angular momentum and therefore represents a luminosity-weighted ratio of ordered motion (i.e. rotation) to random motion. $\lambda_{R_e}$ is not to be confused with the classical, dynamical spin parameter $\lambda$ \citep{peebles1971}, although the former is a good approximator of the latter. Studies have utilised $\lambda_{R_e}$ to better understand the dynamics of early-type galaxies \citep{moranloh2007, emsellem2011, cappellari2016, graham2018} and stellar bars \citep{rawlings2020}, as well as distinguish between different morphologies \citep{vandesande2021}. In the case of Figures~\ref{fig:aio-lb}~and~\ref{fig:aio-lbrd}, we find that $L_{\text{bar}}$ and $L_{\text{bar}}/R_d$ exhibit opposite trends with respect to $\lambda_{R_e}$. In the case of the absolute bar length, we see a steady increase with increasing $\lambda_{R_e}$. However, when normalised with respect to the scale radius of the galaxy, we find that bars are actually longer in low-spin galaxies. This particular result is consistent with the analysis of bars based on the classical spin parameter in \citet{cervantes-sodi2013}, in which low spin galaxies are more prone to self-interacting gravitational instabilities \citep{combes1981, sellwood1993, athanassoula2003}, compared to the high ordered rotation found in galaxies with higher spin parameters. That the trends are reversed also implies that low-$\lambda_{R_e}$ galaxies tend to be physically larger than their high-$\lambda_{R_e}$ counterparts. Indeed, it is known that $\lambda_{R_e}$ depends strongly on morphology, in that early type galaxies tend to have low $\lambda_{R_e}$ values compared to late type galaxies \citep{emsellem2007, vandesande2021}. It is thus possible that disc scale radii $R_d$ may more strongly dependent on the spin parameter $\lambda_{R_e}$ compared to the lengths of bars $L_{\text{bar}}$ by themselves.

On that note, Figures~\ref{fig:aio-lb}~and~\ref{fig:aio-lbrd} show how the bar length varies with morphology. In the NA10 case, we can see that bars are longest in early-type galaxies, specifically lenticular and early spirals. There is a sharp reduction in both $L_{\text{bar}}$ and $L_{\text{bar}}/R_d$ towards later morphological types, before a slight increase in bar lengths for very late spiral galaxies. That bars are shorter in late type galaxies compared to early type galaxies has been well established by previous studies, based on both observations of nearby galaxies and $N$-body simulations \citep{elmegreen1985,combes1993,aguerri2009,erwin2005,erwin2019}. For the SAMI galaxies we see a peak bar length at SAMI T-Types 1.0 and 1.5 (S0 and S0/Early Spiral respectively) with shorter bars in late spirals and E/S0, the latter of which could likely be to the U-Net extracting bulges. One explanation for late type galaxies having shorter bars is due to their higher gas content, which can potentially suppress bar formation and/or slow the rate of growth of the bar \citep{bournaud2005, berentzen2007, athanassoula2013}. The previous study by \citet{hoyle2011} found a similar dependence of bar size with colour, with redder, early-type galaxies hosting longer bars than bluer, late-type galaxies. It is worth noting that the bar size dependence in Figures~\ref{fig:aio-lb} and~\ref{fig:aio-lbrd} also correlates with an increasing bar fraction in redder galaxies \citep{masters2011, skibba2012, vera2016, cervantessodi2017}. However, the recent study by \citet{tawfeek2022} is a notable exception having found a peak bar fraction for more massive, late-type galaxies (although this study did not examine bar size), while \citet{erwin2018} also finds that bars are similarly just as frequent in both gas-rich, bluer galaxies as gas-poor, redder galaxies.

Lastly, it is worth considering the impacts of applying different local thresholds to the predicted bar mask on the overall bar lengths, and how this may impact the previous discussion of physical properties. Recall that in Figure~\ref{fig:barcontours} it was demonstrated that the different thresholds clearly impact the estimated bar length by retaining different regions of the predicted bar mask; larger thresholds result in smaller bar lengths. Figure~\ref{fig:lbthres} shows how different, equally-spaced local thresholds shape the dependence of $L_{\text{bar}}/R_d$ on $\log(M_\star/M_\odot)$. It can be seen that the overall trends are almost identical with only minor variation, and that the only key difference amounts to relatively constant scale factors when going from one threshold to another. That the trends remain so similar suggests that changing the local threshold merely changes the estimated length, and that the predicted bar masks are otherwise evenly smooth and uniform. This further implies that it is possible to calibrate a choice of threshold based on known bar lengths for a given dataset.

\begin{figure}
    \centering
    \includegraphics[scale=0.62]{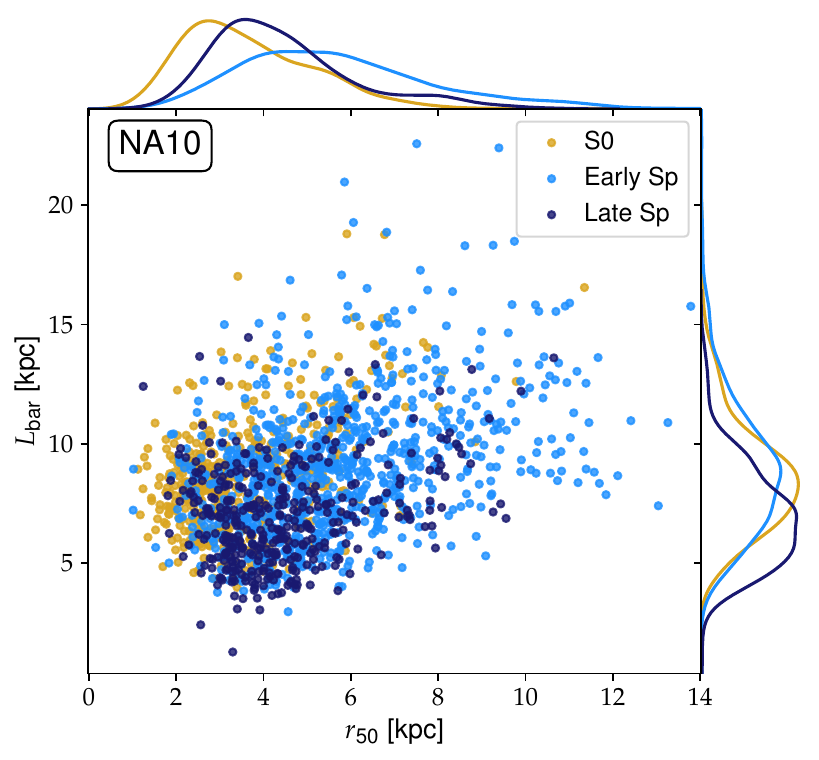}\\
    \includegraphics[scale=0.62]{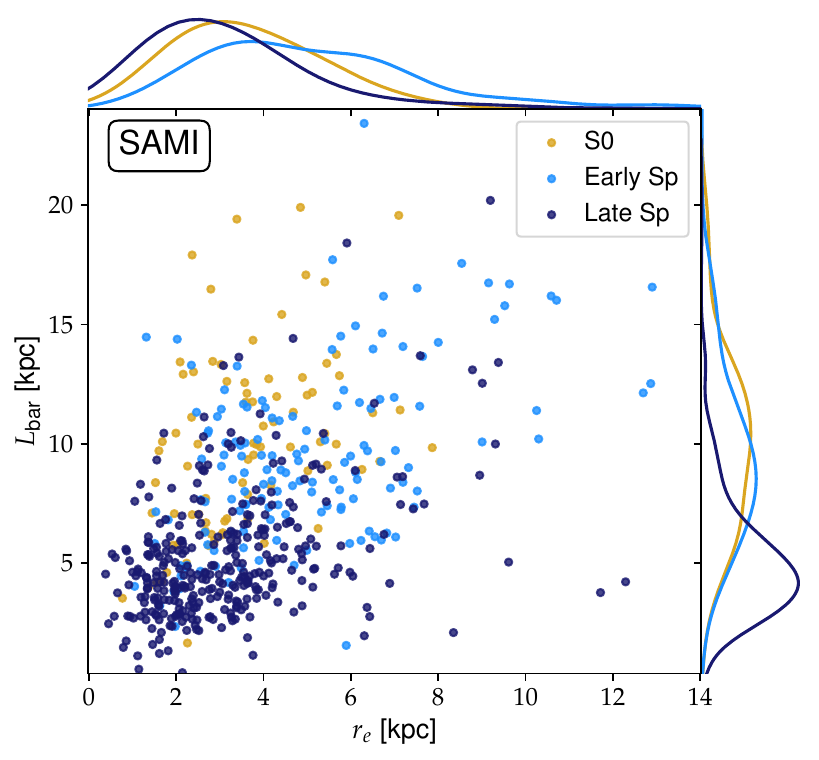}
    \caption{The absolute bar length $L_{\text{bar}}$ in terms of the Petrosian half-flux radius $r_{50}$ for the NA10 galaxies and effective radius $r_e$ for the SAMI galaxies. Points are coloured according to morphology, with the probability densities displayed on the margins.}
    \label{fig:lbar-size}
\end{figure}

\section{Discussion}
\label{s:discussion}

\subsection{Galaxy Size}

\begin{figure}
    \centering
    \includegraphics[scale=0.56]{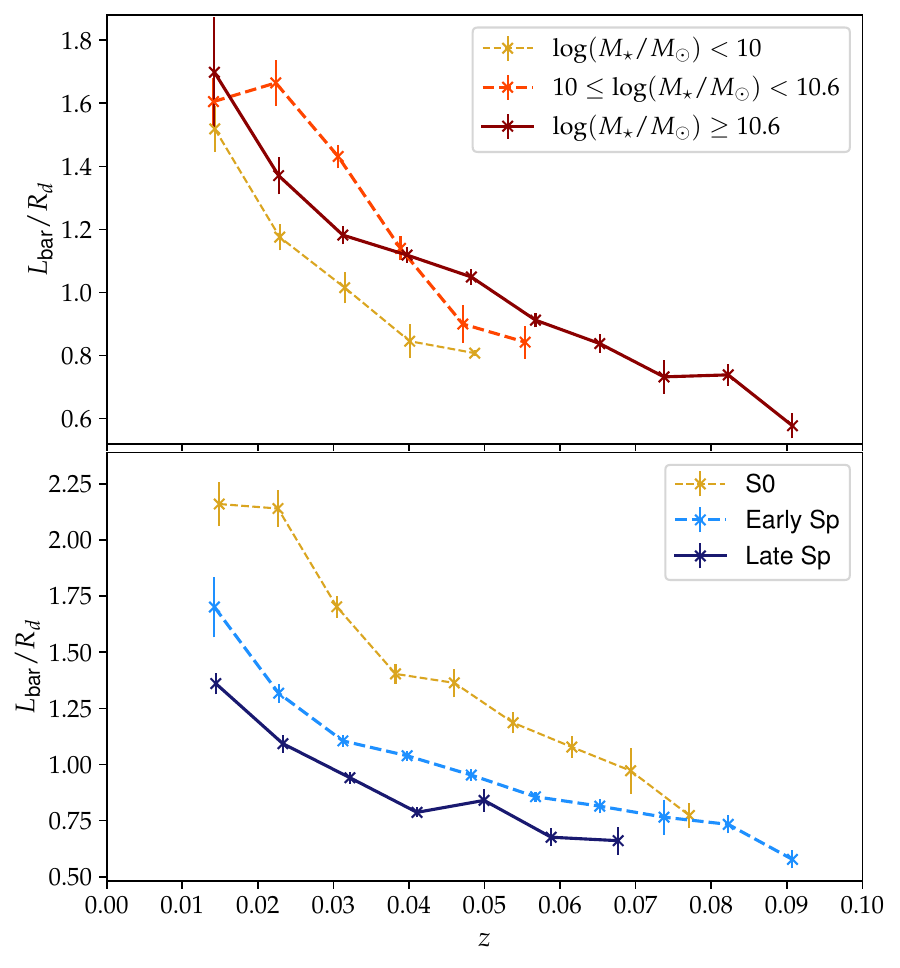}
    \caption{Relative bar length $L_{\text{bar}}/R_d$ as a function of redshift $z$ for NA10 barred galaxies in different mass ranges and for different morphologies.}
    \label{fig:lbar_na10}
\end{figure}
\begin{figure}
    \centering
    \includegraphics[scale=0.57]{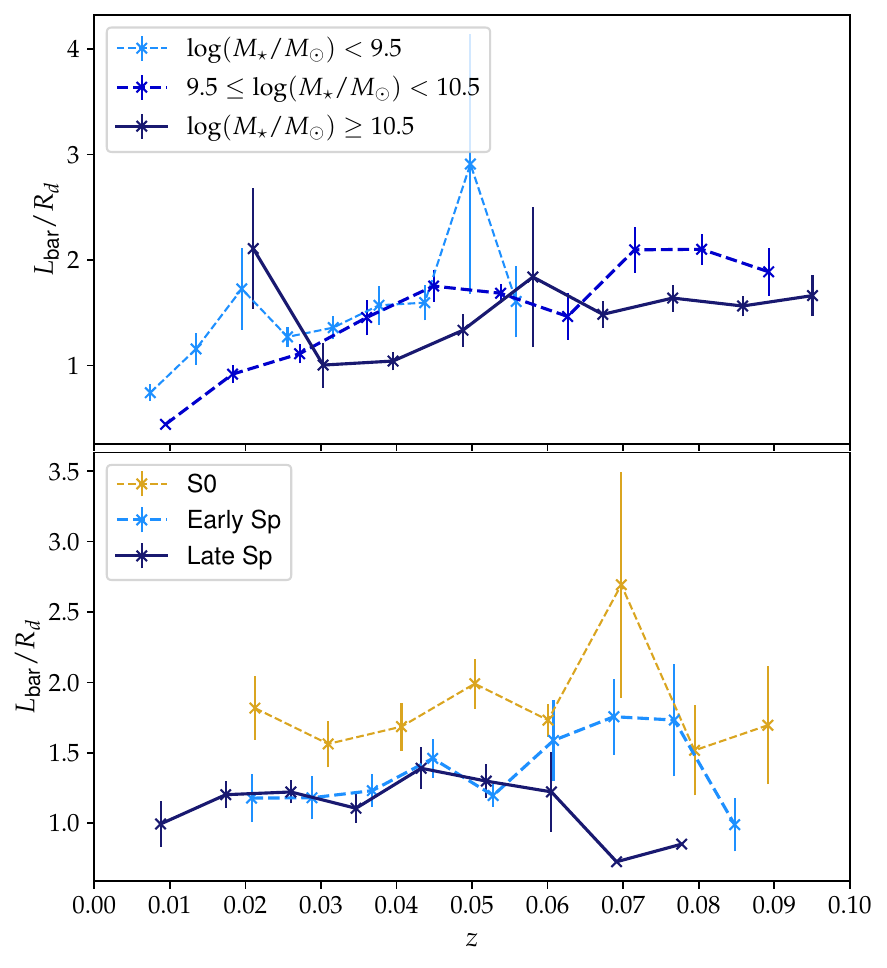}
    \caption{Relative bar length $L_{\text{bar}}/R_d$ as a function of redshift $z$ for SAMI barred galaxies in different mass ranges and for different morphologies.}
    \label{fig:lbar_sami}
\end{figure}

\begin{figure*}
    \centering
    \includegraphics[scale=0.58]{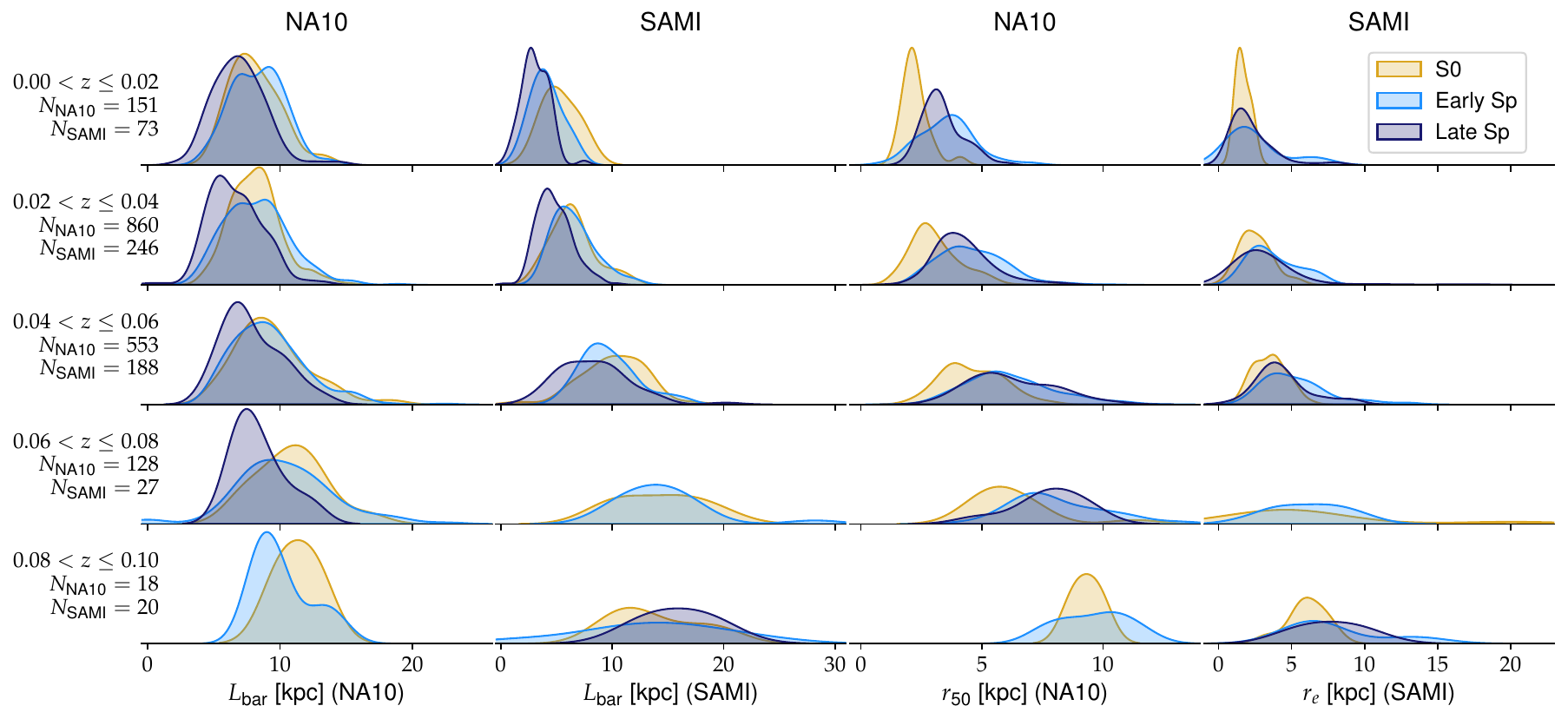}
    \caption{Probability density distributions for the absolute bar sizes $L_{\text{bar}}$ and galaxy sizes $r_{50}$ (NA10) and $r_e$ (SAMI) across different redshift ranges, separated by morphological type.}
    \label{fig:probdensities}
\end{figure*}

So far we have established that the bar length varies with respect to various physical properties including stellar mass and morphology. Although we have examined the normalised, relative bar length $L_{\text{bar}}/R_d$, it is worth examining the physical sizes of galaxies in their own right. Figure~\ref{fig:lbar-size} illustrates the distributions of NA10 and SAMI galaxies, with their absolute estimated bar length $L_{\text{bar}}$ plotted against their physical size, grouped by morphology. Here, early spiral and late spiral designations are grouped by dividing the spiral T-Type morphologies. In the case of NA10, our early spiral group includes T-Types 2 through 4 (Sab to Sbc), with late spiral containing T-Types 5 and above, excluding irregulars and miscellaneous galaxies (Sc to Sm). For SAMI, the division is T-Types 2 and 2.5 (early spiral, early/late spiral) for early spiral, and T-Type 3 (late spiral) for late spirals, excluding indeterminate/unknown types. T-Type 1.5 (S0/early spiral) was assigned to the S0 category in order to better balance the partition. Note also that the definitions are not directly comparable due to the inherent difference between the NA10 and SAMI T-Type definitions \citep{nair2010,cortese2016}.

It can be seen in Figure~\ref{fig:lbar-size} that there is significant scatter when it comes to morphology, however it is still possible to disentangle both the NA10 and SAMI datasets. As expected from Figure~\ref{fig:aio-lb}, the absolute bar length $L_{\text{bar}}$ tends to be shortest in late spirals, although the distributions for $L_{\text{bar}}$ with respect to S0s and early spirals are a lot tighter. The distribution of $L_{\text{bar}}$ is best separated with morphology in the SAMI dataset. When it comes to galaxy size, the smallest galaxies in SAMI tend to be late spirals, while the smallest galaxies in NA10 ($r_{\text{50}} \leq 2$ kpc) are overwhelmingly S0 galaxies. Galaxy size $r_{50}$ acts as fair indicator of morphology, as evinced by their plotted probability distributions, with S0s mostly present at low $r_{50}$, followed by late spirals and then early spirals with increasing $r_{50}$. Curiously, there is a very sharp drop in the number of late spirals with absolute bar lengths $L_{\text{bar}} > 10$ kpc. 

The results of Figure~\ref{fig:lbar-size} also better explain the dramatic differences in trends with $L_{\text{bar}}$ and $L_{\text{bar}}/R_d$ with respect to $\lambda_{R_e}$ that were established in Figures~\ref{fig:aio-lb}~and~\ref{fig:aio-lbrd}; in particular, that low-$\lambda_{R_e}$ galaxies (predominantly early-type) tend to be physically larger than their high-$\lambda_{R_e}$ counterparts (predominantly late-type), at least with respect to the barred galaxies examined in this study. There is yet to be a large-scale, systematic study of the kinematic differences between barred and unbarred galaxies in SAMI, however our results on bar lengths with respect to $\lambda_{R_e}$ demonstrates the importance of ordered rotation in limiting the length of stellar bars with respect to the scale radius of its host galaxy. However, we stress that the results of Figure~\ref{fig:lbar-size} are necessarily dependent on sample selection.

\subsection{Evolution of Bar Length}

Figure~\ref{fig:aio-lb}~and~\ref{fig:aio-lbrd} demonstrated that bar length varies considerably with redshift, with NA10 and SAMI exhibiting opposite trends in the case of the relative length $L_{\text{bar}}/R_d$. In particular, the NA10 galaxies showed a steady decline in relative bar length with increasing $z$, while the SAMI galaxies showed an increase. Of course, the change in galaxy size over redshift is also influenced by sample selection effects, and hence NA10 and SAMI are not directly comparable. However, by examining the changes in bar length over time with respect to other physical properties, such as stellar mass and morphology, it is possible to glean further insights into what could be driving this change.

Figure~\ref{fig:lbar_na10} examines the decline in the relative lengths of bars for barred galaxies divided into different mass ranges and different morphological types. Already it can be seen that NA10 only samples low to intermediate-mass galaxies at relatively low redshifts, sampling only the most massive galaxies at high $z$. Nevertheless, the relative bar length declines in all three mass ranges, with the steepest decline for intermediate-mass samples between $0.02 < z < 0.05$. In terms of morphology, we see that the bar length for S0s exhibits the largest decline, from well over $2R_d$ at low redshift to around $0.8R_d$ at $z \approx 0.08$. In general, the rate of decline in bar length is highest at low redshift, which implies that the growth of stellar bars is accelerating.

\begin{figure*}
    \centering
    \includegraphics[scale=0.4]{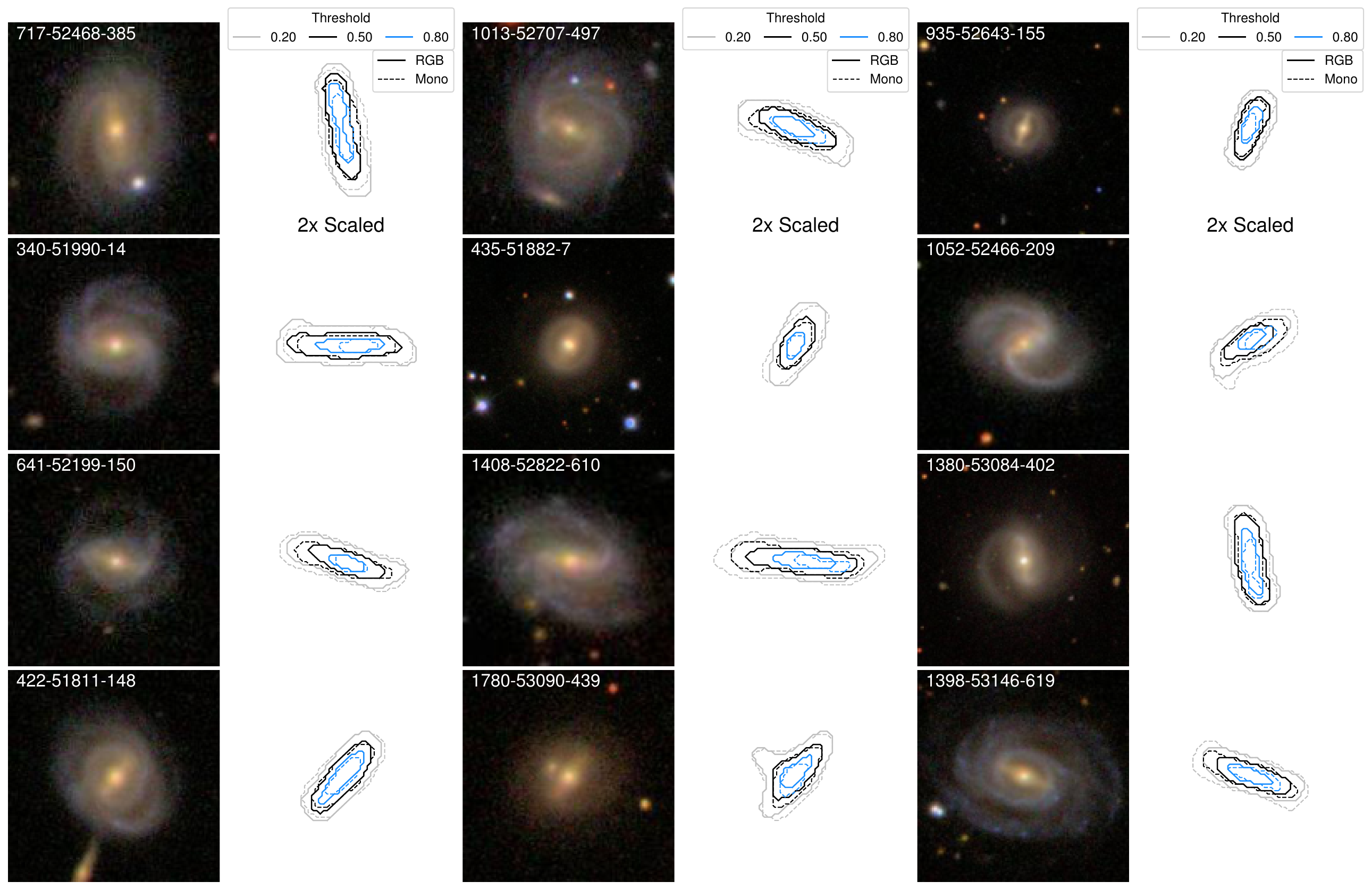}
    \caption{Illustration of the contours for the same galaxies as in Figure~\ref{fig:barcontours}, but this time showing the difference in the contours of the bar mask for the RGB colour U-Net and the monochrome U-Net at a given local threshold.}
    \label{fig:barcontours-rgbvsmono}
\end{figure*}

We saw in the previous section that the SAMI galaxies exhibit different trends compared to the NA10 galaxies. Indeed, Figure~\ref{fig:lbar_sami} shows the increase in relative bar size with increasing redshift. Compared to Figure~\ref{fig:lbar_na10}, there is less overall separation for galaxies of different masses, with the low and intermediate mass galaxies showing similar trends. We note that the mass ranges in Figures~\ref{fig:lbar_na10}~and~\ref{fig:lbar_sami} are slightly different; this is done to account for the different sample selections and to ensure that the three mass range groups each have a similar number of galaxies. Examining morphology, and again lenticular galaxies have longer normalised bar lengths, with both the early and late spiral morphological types having similar bar lengths for $z < 0.05$, beyond which bars in early spirals are longer. However, beyond $z > 0.05$ there are relatively few galaxies in the SAMI sample, hence the greater uncertainty. That $L_{\text{bar}}/R_d$ is highest in S0s compared to the other types may be reflective of higher baryonic fractions in S0s, possible as a result of the gas stripping associated with their formation \citep{bekki2011}. Thus bars which may not have been as prominent in early spirals may be larger relative to $R_d$ in S0s after having undergone a faded spiral morphological transition \citep{rizzo2018, deeley2020, coccato2022}.

\begin{figure*}
    \centering
    \includegraphics[scale=0.565]{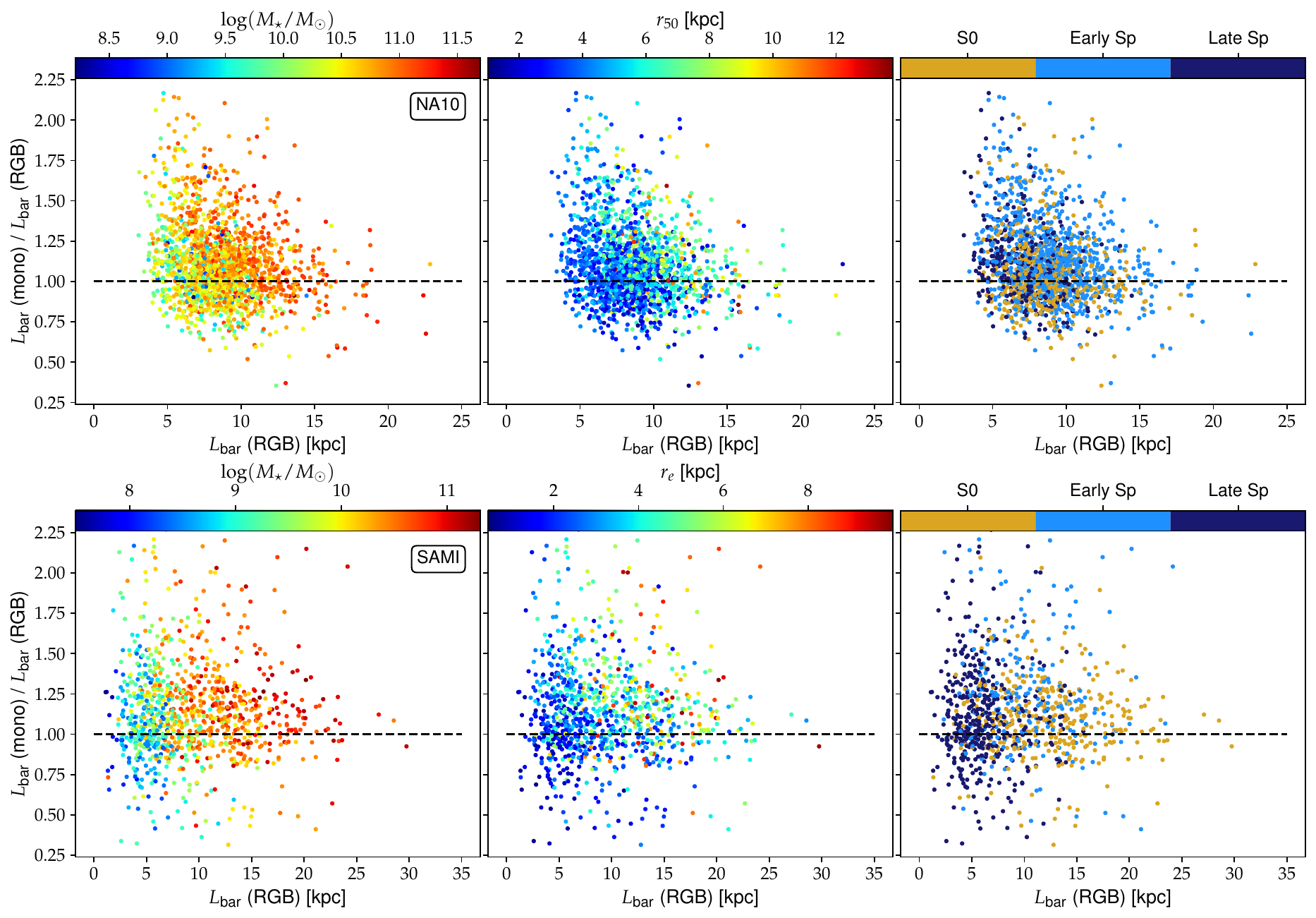}
    \caption{Comparison of the estimated bar lengths $L_{\text{bar}}$ (mono) and $L_{\text{bar}}$ (RGB) obtained from the predicted masks of the monochrome and RGB colour U-Nets respectively. The vertical axis shows the ratio $L_{\text{bar}}$ (mono) / $L_{\text{bar}}$ (RGB), with the black dotted line denoting equal lengths. From left to right, the samples are coloured according to stellar mass, galaxy size and morphology.}
    \label{fig:lengths-monovsrgb}
\end{figure*}

The reduction in $L_{\text{bar}}/R_d$ over time (as observed for the SAMI galaxies) can be interpreted in two ways; that the bars are shrinking over time, or that the host galaxies are shrinking. Given the previously established results in Figure~\ref{fig:aio-lb}, this would suggest that the sizes are shrinking over time. To confirm this, Figure~\ref{fig:probdensities} shows the changes in the distributions of absolute bar lengths $L_{\text{bar}}$ and galaxy physical sizes for both the NA10 and SAMI galaxies over different redshift ranges, separated by morphological type. Indeed, this helps to clarify the results of both Figures~\ref{fig:lbar_na10}~and~\ref{fig:lbar_sami}. In the case of SAMI, there is a large increase in $L_{\text{bar}}$ with increasing $z$, which exceeds the corresponding increase in galaxy physical size $r_e$ over the same $z$ intervals. This is why, for SAMI, $L_{\text{bar}}/R_d$ increases with increasing $z$. However, in the case of NA10, the overall distributions of bar lengths $L_{\text{bar}}$ remains relatively similar, with only a modest growth over time. However, the physical sizes of NA10 galaxies differ dramatically with $z$. This is why, for NA10, $L_{\text{bar}}/R_d$ decreases with increasing $z$.

It should be stressed that the changes in the distributions of galaxy physical sizes are largely a consequence of sample selection, and care should be taken to interpret this as representative of all galaxies with a given morphological type in these redshift ranges. However, this may not wholly account for the dramatic difference in the changes in the distributions of $L_{\text{bar}}$ between the NA10 and SAMI galaxies. In particular, it is curious to note how the distributions of the different morphologies are relatively similar at low redshifts, but become more distinct at higher redshifts in that the peaks of each morphology are more separated. However, this divergence could also be a side effect of low number statistics, as shown by the lack of any late spirals in NA10 beyond $z > 0.08$. Another explanation for the decline in $L_{\text{bar}}/R_d$ with decreasing $z$ for the SAMI galaxies is the fact that spirals account for almost all the SAMI galaxies with $L_{\text{bar}} < 5$ kpc (as per Figure~\ref{fig:lbar-size}) and that, unlike the NA10 galaxies, the spirals in SAMI all peak at the same physical size as lenticulars. Recall from Figure~\ref{fig:aio-lbrd} that spirals have shorter bars than lenticulars.

The results of Figures~\ref{fig:lbar_na10}, \ref{fig:lbar_sami} and \ref{fig:probdensities} suggest that there is evolution in bar lengths, both absolute and normalised, for nearby galaxies with redshifts below $0.1$. However, it is important to contextualise this in terms of the different sample selection criteria employed for the NA10 and SAMI surveys. Both samples are biased towards more massive galaxies, well above $10^{10} M_\odot$, at their highest redshift ranges \citep{nair2010, bryant2015}. This is a key driver of the increase in galaxy sizes with increasing redshift as observed in Figure~\ref{fig:probdensities}, given that these more massive galaxies tend to be physically larger. This selection effects have a key impact on the evolution of $L_{\text{bar}}/R_d$ with $z$. Studies examining galaxies out to higher redshifts have found little to no change in normalised bar length, implying that bars continue to scale with their host galaxies' discs \citep{kim2021}. Indeed, previous studies have established that, as bars evolve over time, they tend to grow longer \citep{athanassoula2003,erwin2005,hoyle2011,diaz-garcia2016,kruk2018}. Furthermore, it is also worth noting that observational studies have found that the bar fraction declines with increasing redshift \citep{sheth2008,cameron2010,melvin2014}, with similar trends also found in simulations \citep{algorry2017,rosas-guevara2019,cavanagh2022}. The bar fraction is known to be greatest in spiral galaxies \citep{eskridge2000,aguerri2009,saha2018}. Indeed, the majority of the known bars in NA10 and predicted bars in SAMI are spirals (see Figure~\ref{fig:lbar-size}).

\subsection{Comparison of the RGB Colour \& Monochromatic U-Nets}
\label{ss:rgbmono}

Although the GZ3D dataset provides RGB colour cutouts, this does not preclude the possibility of training a U-Net to classify greyscale, monochromatic imaging. Such a U-Net could enjoy wider applications to datasets where colour imaging is not readily available, and/or potentially be utilised to examine single bands individually. The purpose of this section is therefore to examine the differences in both the predicted bar masks and estimated bar lengths between the two U-Nets; one which uses RGB colour imaging as its input, the other which uses monochrome imaging. As aforementioned in Section~\ref{ss:unetmodel}, we trained our monochrome U-Net by converting the RGB cutout GZ3D cutouts to greyscale using Python's \textsc{Pillow} package. Of course, we note that it is possible to use imaging from a specific photometric band to train such a monochrome U-Net, upon which the performance in different bands (for instance, $g$-band vs. $i$-band) could be examined, however such investigations are beyond the scope of the current paper. The purpose of this section is to demonstrate that the U-Net model is not strictly restricted to colour imaging, but can also be applied to monochrome imaging. For the sake of simplicity, and to guarantee the correctness of the GZ3D count masks, we have chosen to train the monochrome U-Net model with greyscale versions of three-band colour imaging, but this is without loss of generality and does not preclude using single-band imaging.

\begin{figure*}
    \centering
    \includegraphics[scale=0.52]{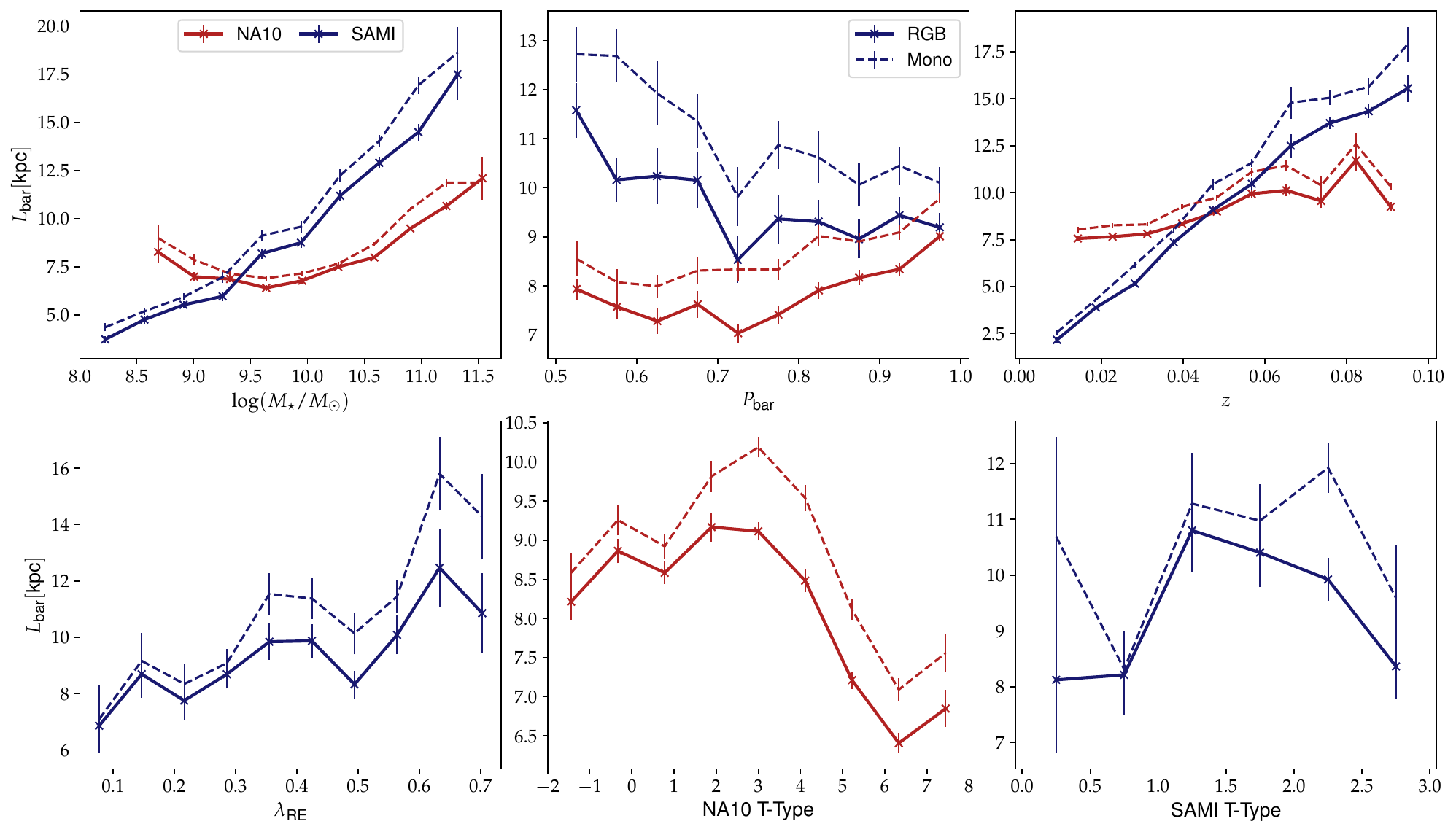}\\
    \includegraphics[scale=0.52]{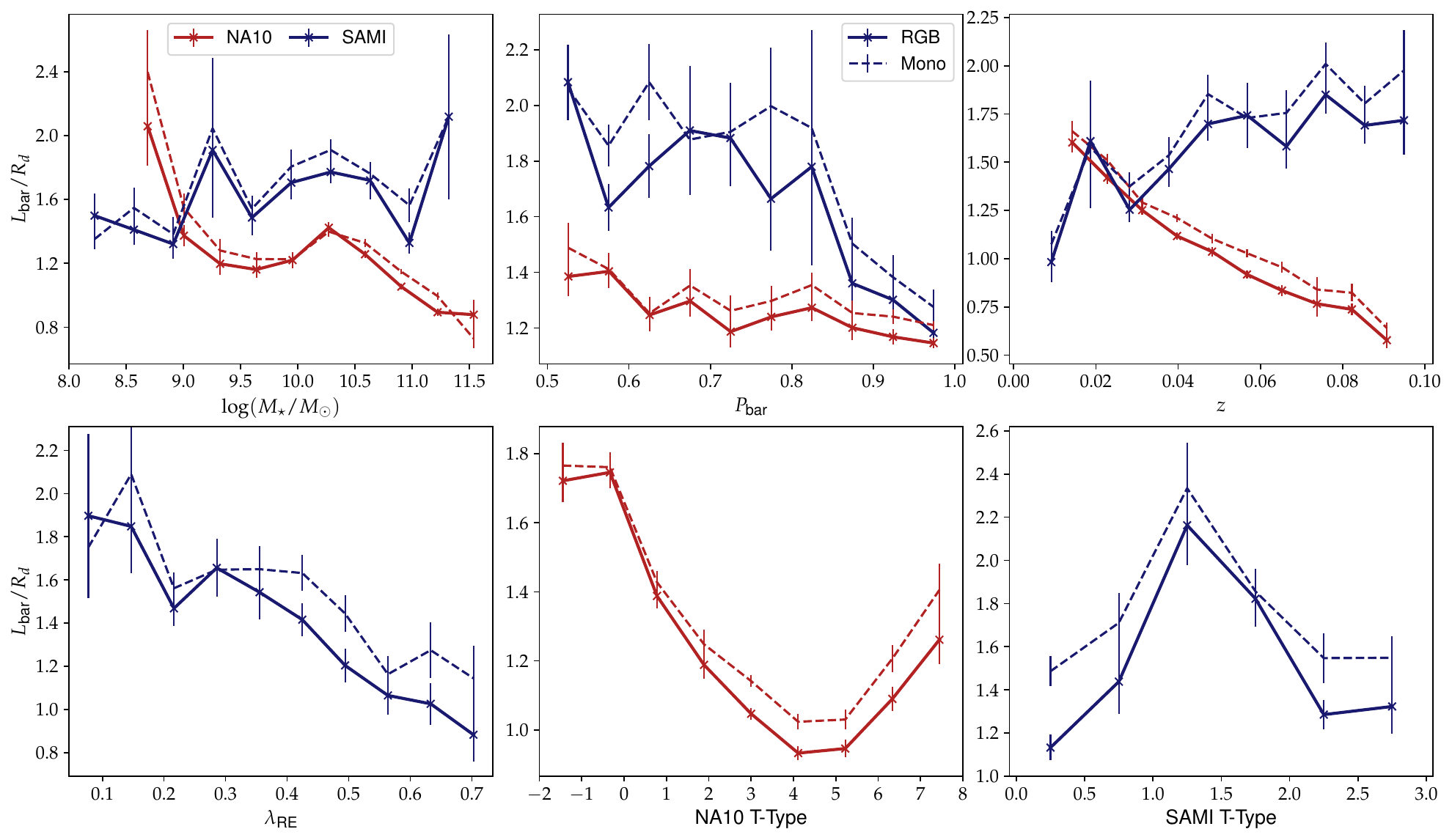}
    \caption{The same plots as in Figures~\ref{fig:aio-lb} and~\ref{fig:aio-lbrd}, but this time showing the results for both the RGB colour and monochrome U-Nets.}
    \label{fig:aio-rgbmono}
\end{figure*}

Figure~\ref{fig:barcontours-rgbvsmono} shows a comparison of the extents of the predicted bar masks from the RGB colour U-Net and monochrome U-Net for three different local thresholds. In general, the contours maintain the same overall width and orientation, but mainly differ in length. At a given threshold, the monochrome predicted mask seems to extend further out than the corresponding colour mask. However, there are some cases in which the reverse is true, particularly at the 80\% threshold. It can also be seen at the 80\% threshold that the predicted masks are offset, and so the central coordinates of the predicted bar differ. We do note that the pixel values for the predicted bar masks are designed to reflect the confidence the given pixel corresponds to a bar feature. The discrepancy could thus follow from the fact that one half of the bar is more easily distinguishable than the other half, shifting the focus of the predicted mask. This also explains why the high-threshold offset appears to affect the monochrome U-Net more than the RGB U-Net; after all, the bar feature is more distinguishable from the rest of the galaxy in a colour image. On a similar note, it can be seen in Figure~\ref{fig:barcontours-rgbvsmono} that there are some cases where the 50\% threshold monochrome contour stretches out as far as the 20\% threshold. This is especially visible in the spiral galaxies 1408-52822-610 and 1398-53146-619. Also, in the spiral galaxy 1052-52466-209, it can also be seen that the monochrome contour begins to trace along the spiral arm. This is likely due of the increased difficulty in differentiating between the end of the bar and the start of the spiral arm resulting from the loss of colour information. This is where training on a specific photometric band, such as the $i$ band where bars are more visually prominent, may be more beneficial. Taken altogether, the results of Figure~\ref{fig:barcontours-rgbvsmono} nevertheless demonstrate that the monochrome U-Net is able to successfully extract bars as distinct, separate features from spiral arms, albeit with a greater degree of uncertainty as illustrated with the wider contours.

It is useful to compare the predicted lengths $L_{\text{bar}}$ from the monochrome U-Net with that from the RGB colour U-Net; this is shown in Figure~\ref{fig:lengths-monovsrgb}, where the points are coloured according to stellar mass, galaxy size, and morphology. In general, there is a fairly high degree of scatter, but in the majority of cases the predicted bar length from the monochrome model is greater than that from the colour model. Figure~\ref{fig:lengths-monovsrgb} also appears to show that the discrepancies are higher for smaller, less massive galaxies where $L_{\text{bar}}$ (RGB) is itself smaller. Figure~\ref{fig:lengths-monovsrgb} also confirms the clear differences between the NA10 and SAMI samples that was observed in Figures~\ref{fig:probdensities}, especially regarding morphology where the smallest SAMI galaxies are almost all predominantly late spirals, compared to a more even distribution across morphologies for the NA10 samples. In the case where $L_{\text{bar}}$ (mono) $ > 1.5 \times L_{\text{bar}}$ (RGB), the majority of samples are spirals. This is to be expected given the inherent uncertainty in distinguishing the ends of bars from the starts of spiral arms, as hinted in Figure~\ref{fig:barcontours-rgbvsmono}.

Lastly, it is worth inspecting how the use of the monochromatic U-Net alters the trends in $L_{\text{bar}}$ with respect to physical properties as previously investigated in Section~\ref{s:results}. Figure~\ref{fig:aio-rgbmono} presents the results of Figures~\ref{fig:aio-lb} and~\ref{fig:aio-lbrd}, this time including the corresponding trends from the monochrome model. As expected from Figure~\ref{fig:lengths-monovsrgb}, the lengths from the monochrome U-Net are generally higher than the corresponding lengths from the colour U-Net. However, the differences between the two lengths appear to vary with respect to several properties. Good examples of this are the discrepancies in both $L_{\text{bar}}$ and $L_{\text{bar}}/R_d$ with the NA10 T-Type and $\lambda_{R_e}$. Furthermore, Figure~\ref{fig:aio-rgbmono} appears to support the aforementioned greater uncertainty in distinguishing bars and spiral in monochromatic imaging. In the case of the NA10 dataset, there is a higher discrepancy for later morphological types compared to early types, with a similar effect seen with the SAMI galaxies, especially in the higher discrepancy for high-$\lambda_{R_e}$ galaxies which are predominantly spirals.

\subsection{Model Benefits and Limitations}
\label{ss:modeldisc}

Our model developed in this work offers a unique approach to analysing and studying bars in barred galaxies. However, while this approach has some key advantages and benefits, there are also some crucial limitations. The major advantage of our image segmentation approach is that our U-Net directly outputs spiral and bar masks for a given image of a galaxy without the need for auxiliary data or free parameters / thresholds that must be given initial values, such as in isophotal ellipse fitting. Furthermore, the U-Net can make these predictions extremely quickly, obtaining masks for thousands of galaxies in a matter of seconds, as previously stated in Section~\ref{s:methods}. In tandem with our bar model CNN used to classify barred galaxies from \citet{cavanagh2022}, this enables our U-Net to be readily applied to study bar lengths in large-scale observational surveys and even cosmological simulations. 

Another benefit is the manner in which the U-Net is configured with respect to the training data; namely for pixel-level regression rather than pixel-level classification. The decision for our U-Nets to directly output a smooth, continuous mask reproducing the GZ3D count masks is mostly intended to simplify and expand downstream applications. As aforementioned in Section~\ref{s:results}, it is possible to calibrate a suitable local threshold for a given dataset of barred galaxies based on known bar lengths. This is an important advantage of our approach; it would not be possible were the U-Net instead pre-trained on masks with a fixed, pre-determined threshold, for all future applications would be subject to that hard-coded threshold. The estimated bar lengths are ultimately dependent on a choice of threshold, however it can be argued that this is an inevitable necessity to estimate bar lengths, whether from pixel masks or isophotal ellipse fitting. Since our aim is to reproduce the GZ3D count masks, we have kept the original count values untouched, having simply rescaled each pixel value by dividing by the maximum number of volunteers. However, this does not rule out the use of other nonlinear or weighted scaling methods to better standardise these counts, such as by minimising the impact of galaxies with high numbers volunteer counts, or instead using count fractions to instead formulate the masks as probabilities, in which case the U-Net could be configured for pixel-level classification with a Softmax activation. These refinements are a focus of future experimentation, and we stress that the U-Net models presented in this paper are not restricted to any one particular method of count mask scaling or preprocessing.

Given that the GZ3D count masks constitute the entirety of the training data used to train the U-Net, they ultimately impose important fundamental limitations on our U-Nets. Errors and biases in these masks will necessarily filter through into the U-Net, and be reflected in the U-Net's predictions. This has already been glimpsed in the complexity of the predicted spiral masks, which is to be expected given the more diverse range of possible spiral arm arrangements compared to bars. We will discuss the performance of the model when subject to corrupted count masks in more detail in Appendix~\ref{app:inputs}. Furthermore, as illustrated in Figure~\ref{fig:gz3dexample}, the image cutouts provided as part of the GZ3D data release all feature a purple hexagon, which is less than ideal from the perspective of pure image segmentation. Fortunately, its presence does not appear to have any adverse effect on the predicted bar and spiral masks. We also note that we did not utilise data augmentation techniques when training the U-Net. Here, data augmentation is a general term referred to artificially increasing the size of the training data through modifying existing images, most often by applying affine transformations such as random rotations, mirroring or scaling, or even cropping and translation. Of course, any augmentation would have to be applied equally to both the input image (GZ3D cutout) and the target spiral and bar count masks, which has the potential to introduce artefacts. As such, to ensure the closest possible reproduction of the count masks, and for the sake of computational simplicity, we elected not to utilise data augmentation.

One of the difficulties in detecting bars is poor image quality \citep{kim2016}. While our study is limited to nearby galaxies, and therefore does not probe deep enough to encounter significant resolution limits, low image quality nevertheless has an impact on the ability of the U-Net to extract suitably uniform bar masks, ultimately affecting the estimated bar lengths at a given threshold. This therefore places an upper limit on redshifts at which our current model can be feasibly applied, keeping in mind other considerations such as $k$-correction and evolution. Such high-redshift applications will likely require the use of transfer learning techniques in order to adapt the current U-Net to account for these differences (see \citealt{dominguezsanchez2019,cavanagh2023}).

One additional application for our U-Net is examining bar evolution in cosmological simulations, where mock RGB imaging can be readily created at various snapshots in redshift. Studies have long utilised simulations to study bars, from the pioneering $N$-body simulations of \citet{hohl1971,combes1981,pfenniger1984,miwa1998}, to more recent studies examining bars in large-scale hydrodynamical simulations \citep{algorry2017,zhou2020,roshan2021,cavanagh2022,rosas-guevara2022}. Given that our model utilises an image segmentation approach, it is especially well suited to the finer resolution of zoom-in simulations, where the physical processes affecting bars can be studied in greater detail \citep{martig2012,zana2019}. Another promising application is using the predicted U-Net bar masks not for estimating bar lengths, but instead as an overlay atop a mass or photometric flux map in order to estimate the mass fraction of the bar.

\section{Conclusions}
\label{s:concl}

Our main results are summarised below:
\begin{enumerate}[(i)]
\item We have developed a deep learning model to perform morphological segmentation of galaxy imaging in order to estimate the lengths of bars in barred galaxies and examine how both the absolute bar length $L_{\text{bar}}$ and normalised bar length $L_{\text{bar}}/R_d$ varies with respect to various physical properties. This represents a novel and inherently versatile approach for extracting bar masks using solely the image of a galaxy, without any auxiliary information. We have demonstrated the versatility and efficiency of this new method through its application to two different datasets with different source imaging, as well as to both RGB and monochrome imaging. In particular, we classified known bars from the NA10 morphological catalogue, as well as predicted barred galaxies from the SAMI catalogue as classified with a CNN from our previous work.
\item We have found that, in terms of absolute length $L_{\text{bar}}$ bars in high-mass galaxies are physically longer than in low-mass galaxies. However, this is not necessarily reflected when examining the normalised bar length. Our results for the NA10 dataset demonstrated a strong decline in $L_{\text{bar}}/R_d$ with increasing stellar mass, while the SAMI dataset demonstrated mixed results with a weak overall increase.
\item We have found that bar length also depends strongly on morphology. In general, bars in early-type galaxies are longer than bars in late-type galaxies. We have established a similar result when further partitioning the spiral galaxies, namely that bars in early-type spirals are longer than bars in late-type spirals.
\item We have found that bars in low spin parameter $\lambda_{R_e}$ galaxies are longer, with respect to their host galaxy, than in galaxies with high spin parameters. This is likely also reflective of the morphology of the host galaxy, given that low-$\lambda_{R_e}$ galaxies tend to primarily be early type galaxies (and vice versa).
\item We have found that the distribution of bar lengths evolves with redshift. In the case of the NA10 galaxies, we have found a strong decrease in $L_{\text{bar}}/R_d$ with increasing $z$ with only a modest increase in $L_{\text{bar}}$. We have also found that the rate of change varies with morphology, with S0s exhibiting the strongest rate of decline. In the case of the SAMI galaxies. We note that these trends are likely strongly driven by the changes in galaxy size and therefore ultimately dependent on sample selection.
\item We have shown that our U-Net model is able to successfully differentiate between spiral arms and stellar bars in monochromatic imaging. However, we found that the predicted bar masks in monochromatic imaging tend to be larger than in colour imaging, subsequently leading to a larger predicted bar length. This is reflective of the greater uncertainty in establishing the boundary between stellar bars and spiral arms in the absence of colour. We further note that $L_{\text{bar}}$ (mono) appears to scale linearly with $L_{\text{bar}}$ (RGB), implying that the two values can be reconciled by applying some scale factor conversion.
\item We note that our U-Net morphological segmentation technique is inherently versatile, and it is possible for the model to be applied to a wider range of observed galaxies. Importantly, our U-Net model can be readily applied to simulated galaxies, where they can potentially play a crucial role in analysing bars and subsequently examining the physical processes governing the formation and growth of bars over cosmic timescales. Apart from just examining the lengths of bars, the predicted bar masks could also be used to estimate the mass fractions of bars.
\end{enumerate}

\section*{Acknowledgements}

This research was supported by the Australian government through the Australian Research Council's Discovery Projects funding scheme (DP220101863). This study heavily utilised the following Python packages and libraries: \textsc{numpy} \citep{harris2020}, \textsc{matplotlib} \citep{hunter2007}, \textsc{seaborn} \citep{waskom2021}, \textsc{Pillow} \citep{vankemenade2022}, \textsc{astropy} \citep{theastropycollaboration2013}, \textsc{TensorFlow} \citep{abadi2016}, \textsc{Keras} \citep{chollet2015} and \textsc{scikit-image} \citep{vanderwalt2014}.

\section*{Data Availability}

This study utilised publicly available data from the NA10, SDSS DR17, Galaxy Zoo 3D, SAMI DR3 and HSC DR3 datasets \citep{nair2010, bryant2015, masters2021, croom2021, abdurrouf2022, aihara2022}. HSC imaging courtesy of the NAOJ/HSC collaboration. Specific data pertinent to the paper will be provided upon reasonable request to the author.

\bibliographystyle{mnras}
\bibliography{mybib_final}

\appendix

\section{Model Performance on Corrupted Inputs}
\label{app:inputs}

\begin{figure*}
    \includegraphics[scale=0.88]{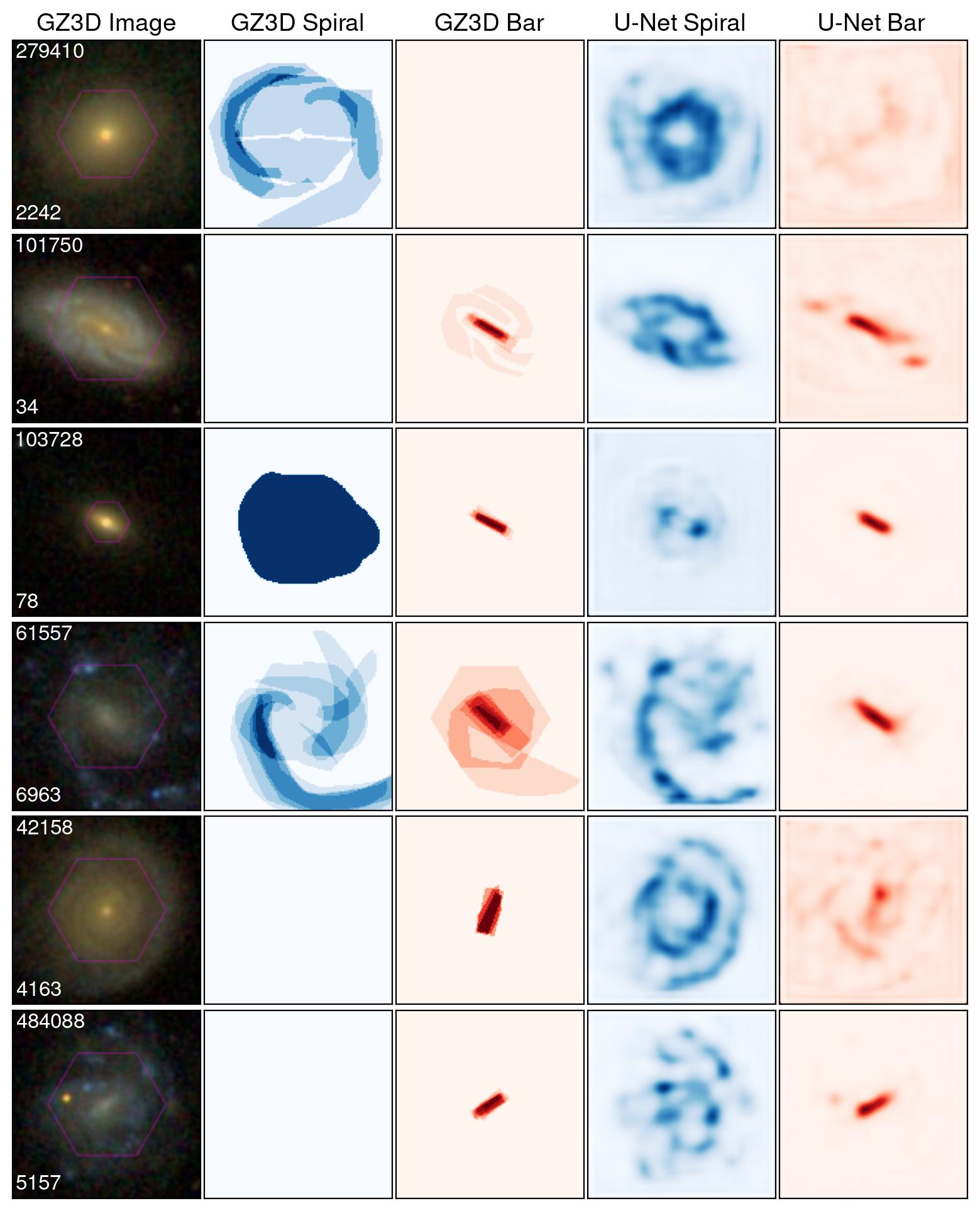}
    \caption{Example application of the U-Net to six randomly selected galaxies from the GZ3D test set with corrupt or missing masks. As with Figure~\ref{fig:gz3dexample}, each row shows the input GZ3D colour cutout, the volunteer-drawn GZ3D spiral and bar masks, and the subsequent predicted spiral and bar masks as directly outputted by the U-Net. The image cutouts are annotated with their GZ3D ID in the top-left.}
    \label{fig:gz3dcorrupted}
\end{figure*}

The GZ3D dataset \citep{masters2021}. Apart from the selection criteria discussed in Section~\ref{s:methods}, namely the selection of galaxies with spiral or bar masks that have been annotated by at least three volunteers, there is no further filtering or cleaning of the training masks. It is important to note that the publicly available GZ3D masks have not been cleaned or otherwise assessed for quality, and there are invariably unrealistic or otherwise corrupted masks present throughout the full data release. Despite applying the minimum volunteer annotation threshold, the training data for our U-Nets likewise includes these corrupted masks. While this does impact the performance of the model, it may actually be beneficial for improving the robustness and reliability of the model. Deliberately corrupting inputs is a known technique in deep learning, used for instance to improve model regularisation and prevent overfitting \citep{jin2016}. These corrupted masks are therefore useful for assessing the overall performance and generalisability of the U-Net.

Figure~\ref{fig:gz3dcorrupted} shows a selection of galaxies from our test set that feature corrupted, missing or otherwise non-ideal GZ3D spiral and bar masks. Despite these anomalous masks, the predicted masks obtained from the U-Net are somewhat sensible, in some cases able to recover or fill-in features that were missing from the GZ3D masks. Examples 101750, 42158 and 484088 show spiral galaxies that have no annotated GZ3D spiral masks. Despite this, the U-Net predicts spiral regions for each of these images. We note that, for these specific examples, the difference between the GZ3D and predicted masks does increase validation loss of the model. This is undesirable from the point of view of the model training, yet the actual outcome is desirable in that the model has filled in the missing spiral masks. Similarly, the galaxy 42158 is unbarred despite being indicated as such in the GZ3D bar mask. The U-Net does not extract a clear bar region, which is indeed desirable since there is no physically-present bar for it to extract. Another noteworthy example is galaxy 61557, which features an erroneous GZ3D bar mask, with some classifiers incorrectly tracing the spiral arm. In this case the U-Net manages to ignore the spiral and correctly output a physically sensible bar mask. When it comes to maximising performance, our U-Net could benefit from greater quality control with regard to the GZ3D training data. This is a focus of future work.

\bsp	
\label{lastpage}
\end{document}